\documentclass[journal,10pt]{IEEEtran}
\usepackage{amsmath,amsfonts}
\usepackage{algorithmic}
\usepackage{algorithm}
\usepackage{array}
\usepackage{subfigure}
\usepackage{textcomp}
\usepackage{stfloats}
\usepackage{url}
\usepackage{verbatim}
\usepackage{graphicx}
\usepackage{cite}
\usepackage[table, dvipsnames]{xcolor}
\usepackage{booktabs,siunitx}
\usepackage{hyperref}
\usepackage{diagbox}
\usepackage{comment}
\usepackage{multirow}
\usepackage{tablefootnote}
\usepackage{enumitem}

\newcommand{\textoverline}[1]{$\overline{\mbox{#1}}$}

\begin{document}

\bstctlcite{IEEEexample:BSTcontrol}

\title{DiaPer: End-to-End Neural Diarization with Perceiver-Based Attractors}

\author{Federico Landini, Mireia Diez, Themos Stafylakis, Lukáš Burget
\thanks{Federico Landini, Mireia Diez, and Lukáš Burget are with Brno University of Technology and Themos Stafylakis is with Omilia and Athens University of Economics and Business.}}

\maketitle

\begin{abstract}
Until recently, the field of speaker diarization was dominated by cascaded systems. Due to their limitations, mainly regarding overlapped speech and cumbersome pipelines, end-to-end models have gained great popularity lately. One of the most successful models is end-to-end neural diarization with encoder-decoder based attractors (EEND-EDA). 
In this work, we replace the EDA module with a Perceiver-based one and show its advantages over EEND-EDA; namely obtaining better performance on the largely studied Callhome dataset, finding the quantity of speakers in a conversation more accurately, and faster inference time. Furthermore, when exhaustively compared with other methods, our model, DiaPer, reaches remarkable performance with a very lightweight design.
Besides, we perform comparisons with other works and a cascaded baseline across more than ten public wide-band datasets.
Together with this publication, we release the code of DiaPer as well as models trained on public and free data.
\end{abstract}


\section{Introduction}
\IEEEPARstart{I}{n} the last years, there has been a big change of paradigm in the world of speaker diarization. Competitive systems until a few years ago were cascaded or modular~\cite{sell2018diarization,landini2020but,park2019auto}, consisting of different sub-modules to handle voice/speech activity detection (VAD/SAD), embedding extraction (usually x-vector) over uniform segmentation, clustering, optional resegmentation and overlapped speech detection (OSD) and handling. The main disadvantages of this framework are that each sub-module is trained independently and optimized for different objectives and that the full pipeline is complex since a few steps need to be applied sequentially, propagating errors from one step to the next one. Furthermore, OSD performance is usually not satisfactory, resulting in high overlap-related errors in cascaded systems.

Since the appearance of end-to-end models, the ecosystem has changed substantially with new approaches constantly appearing~\cite{park2022review}. Neural-based diarization models can be separated into different categories: single-stage systems, which comprise only one model, and two-stage systems, which have two steps where one is a variant of end-to-end model and the other is either based on clustering or on another model. Single-stage systems, such as end-to-end neural diarization (EEND)~\cite{fujita19_interspeech}, where diarization is modeled as per-speaker per-frame binary classification, are trained directly for the task. While the training can be done in different steps (training with 2-speaker simulated data, then adapting to data with variable number of speakers and finally fine-tuning to in-domain data), the inference is performed in a single stage. These methods face difficulties in recordings with several speakers~\cite{horiguchi2022encoder}.
Two-stage systems can be separated into different classes. Models such as target speaker voice activity detection~\cite{medennikov2020target} are trained in an end-to-end manner but make use of an initialization provided by an existing (usually cascaded) model which has to be run priorly at inference time. Other two-stage systems run EEND on short segments (where few speakers are expected) and then perform clustering to join the decisions on short segments. They are known as EEND vector clustering (EEND-VC) and different variants have been proposed~\cite{kinoshita2021integrating,delcroix23_interspeech,bredin23_interspeech}. These approaches present advantages in dealing with several speakers (potentially an unlimited number of them) while having an edge over clustering-based methods on dealing with overlapped speech segments as EEND models usually do.
This categorization is, however, not strict. Some systems do not exactly qualify as ``single'' or ``two'' stage as they have a single stage but include some iterative procedure~\cite{zeghidour2021dive,chen23n_interspeech}.

The simplicity of single-stage EEND systems (where diarization is modeled as per-speaker per-frame binary classification) has brought more attention to them and several variations have been proposed based on this framework. The two main extensions are self-attention EEND (SA-EEND)~\cite{fujita2019end} (where BiLSTM layers are replaced by SA ones) and EEND with encoder-decoder attractors (EEND-EDA)~\cite{horiguchi20_interspeech} (which enables handling variable numbers of speakers), but several others have been proposed:
some of them have been designed for the online scenario~\cite{han2021bw,xue2021online} or making use of multiple microphones~\cite{horiguchi2022multi,horiguchi2023mutual}. The Conformer architecture~\cite{gulati20_interspeech} was used to replace the self-attention layers of SA-EEND in \cite{liu21j_interspeech} and of EEND-EDA in \cite{leung21_interspeech}. 

The Perceiver~\cite{jaegle2021perceiver} is a Transformer~\cite{vaswani2017attention} variant that employs cross-attention to project the variable-size input onto a fixed-size set of latent representations. These latents are transformed by iterative self-attention and cross-attention blocks. By encoding the variable-size input into the fixed-size latent space, the Perceiver reduces the quadratic complexity of the Transformer to linear. In this work, we utilize the Perceiver framework to encode speaker information into the latent space and then derive attractors from them. Using Perceivers allows us to handle a variable number of speakers per conversation while addressing some of the limitations of EDA with a fully non-autoregressive (and iteration-free) scheme. Moreover, we evaluate our model, \textit{DiaPer}, on a wide variety of scenarios. The contributions of our work are:
\begin{itemize}[leftmargin=0.4cm]
    \item Replacement of encoder-decoder structure in EEND-EDA by a Perceiver-based decoder.
    \item Analysis of DiaPer's performance under different architectural choices.
    \item Thorough comparison with EEND-EDA to show DiaPer's improvements.
    \item Proposed architecture that is more lightweight and efficient at inference time, yet performs better than EEND-EDA.
    \item Exhaustive comparison with other works on several corpora.
    \item Clustering-based baseline (including VAD and OSD + overlap handling) results on a variety of datasets and built with public tools.
    \item Release of models trained on free publicly available data.
    \item Public code: \url{https://github.com/BUTSpeechFIT/DiaPer}.
\end{itemize}

\section{Related Works}
Among the EEND variants that are capable of dealing with multiple speakers the most standard one is still EEND-EDA~\cite{horiguchi20_interspeech}. This approach employs long short-term memory (LSTM) layers for encoding frame embeddings and decoding attractors that represent the speakers in the conversation. However, one of the limitations of this approach is the LSTM-based encoder-decoder mechanism itself. In practice, the frame-by-frame embeddings fed to the LSTM encoder are shuffled, clearly removing the time information, and hindering the capabilities of this approach. This is done due to the difficulties LSTMs have to ``remember'' speakers appearing at the beginning of the conversation, especially when processing long sequences. In \cite{pan2022towards}, an alternative is proposed where the input of the LSTM encoder is not shuffled and the LSTM decoder incorporates an attention mechanism. Instead of using zero vectors as input for the decoder, the input is obtained as a weighted sum of the encoder outputs, providing the decoder with better cues. A similar idea is explored in \cite{broughton23_interspeech} where the decoder is fed with summary representations calculated together with embeddings produced by the frame encoder.

Some works have explored non-autoregressive approaches for obtaining attractors with attention-based schemes. 
The first of these works replaces the LSTM-based encoder-decoder with two layers of cross-attention decoder~\cite{rybicka2022end}. In this configuration, the attractors are transformed using the frame embeddings as keys and values and the input attractors, used as queries in the decoder, are obtained as the weighted average of the frame embeddings using their predicted posterior activities as weights. 
However, a set of initial attractors has to be fed into the decoder before an initial set of predictions is produced.
The initial attractors are given by running k-means clustering on the frame embeddings and clustering to the number of speakers in the recording. It is shown that this method can improve by running a few refinement iterations. 

In \cite{fujita2023neural}, the LSTM-based encoder-decoder is also replaced by a cross-attention decoder; however, the set of initial queries that are transformed into attractors is not defined by the output of the model but they are learnable parameters.
The methods in~\cite{rybicka2022end,fujita2023neural} have only shown their capabilities in the two-speaker scenario where the number of speakers is known and where the architecture can be crafted to handle that specific quantity. The extension to more speakers is definitely possible but follow-up works have not yet been published. 

A combination of the aforementioned works is utilized in~\cite{hao2023nn,chen23n_interspeech}. In~\cite{hao2023nn}, in the context of SA-EEND for two speakers, the initial diarization outputs are used to estimate initial attractors and they are refined iteratively with cross-attention decoders with a fixed set of queries (one for each of the speakers) attending to frame embeddings. In~\cite{chen23n_interspeech}, the LSTM-based encoder-decoder is also replaced by layers of cross-attention decoder and three of the initial queries are fixed (but learned during training) and represent ``silence'', ``single speaker'' and ``overlap'' while the other $S$ queries represent each of the speakers in the recording. In the first pass, only the fixed queries are used and then the initial speaker queries are estimated from the frame embeddings, using the average of carefully selected frames given the predicted posterior activities. The set of $S+3$ attractors is refined through a few cross-attention layers in order to produce the final attractors used to obtain the speech activity posteriors. It should be noted that the inference procedure with this method is more complicated than in the original EEND-EDA due to the iterative procedure to estimate first silence, single speaker and overlap attractors and then each of the speakers iteratively.

In~\cite{chen23n_interspeech}, and more recently in~\cite{chen2024attention} (which is concurrent to this work), results are presented with a flexible quantity of speakers but the model relies on an autoregressive scheme since the speakers are iteratively decoded in a second step.

All these approaches present similarities with a more generic architecture: the Perceiver~\cite{jaegle2021perceiver} which iteratively refines a set of latents (queries in cross-attention) informed by an input sequence (keys and values in cross-attention) but in a complete non-autoregressive framework. 

The model we propose in this work generalizes some of the ideas described above and directly tackles the problem of handling several speakers using Perceivers to obtain attractors in an EEND-based framework. We name this approach DiaPer: end-to-end neural diarization with Perceiver-based attractors.

\section{The Model}

\begin{figure}[!t]
\centering
\includegraphics[width=\linewidth]{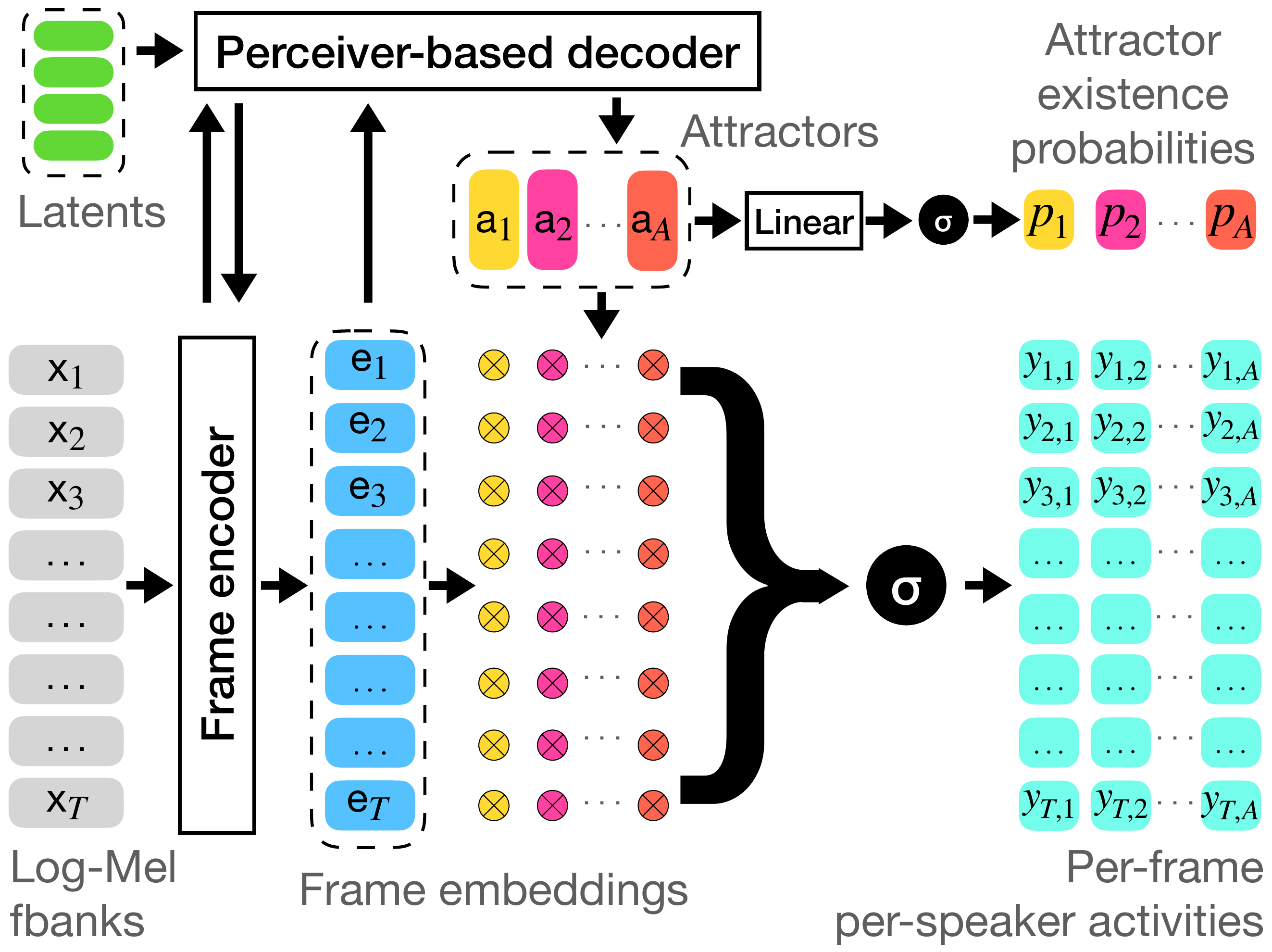}
\vspace{-6mm}
\caption{DiaPer diagram. $\sigma$ refers to the sigmoid function and the circles with crosses mean dot-product between the vectors.}
\label{fig:DiaPer}
\end{figure}

DiaPer shares many facets with other EEND models, such as defining diarization as a per-speaker-per-time-frame binary classification problem. 
Given a sequence of observations (features) $\textbf{X} \in \mathbb{R}^{T \times F}$ where $T$ denotes the sequence length and $F$ the feature dimensionality, the model produces $\hat{\textbf{Y}} \in (0, 1)^{T \times S}$ which represent the speech activity probabilities of the $S$ speakers for each time-frame. Just like with EEND-EDA, the model is trained so that $\hat{\textbf{Y}}$ matches to the reference labels $\textbf{Y} \in \{0, 1\}^{T \times S}$ where $y_{t,s} = 1$ if speaker $s$ is active at time $t$ and silent otherwise. The main difference between EEND-EDA and DiaPer is in how the attractors are obtained given the frame embeddings. As shown in Figure~\ref{fig:DiaPer}, DiaPer makes use of Perceivers to obtain the attractors instead of the LSTM-based encoder-decoder.

\begin{figure*}[!t]
\centering
\includegraphics[width=\textwidth]{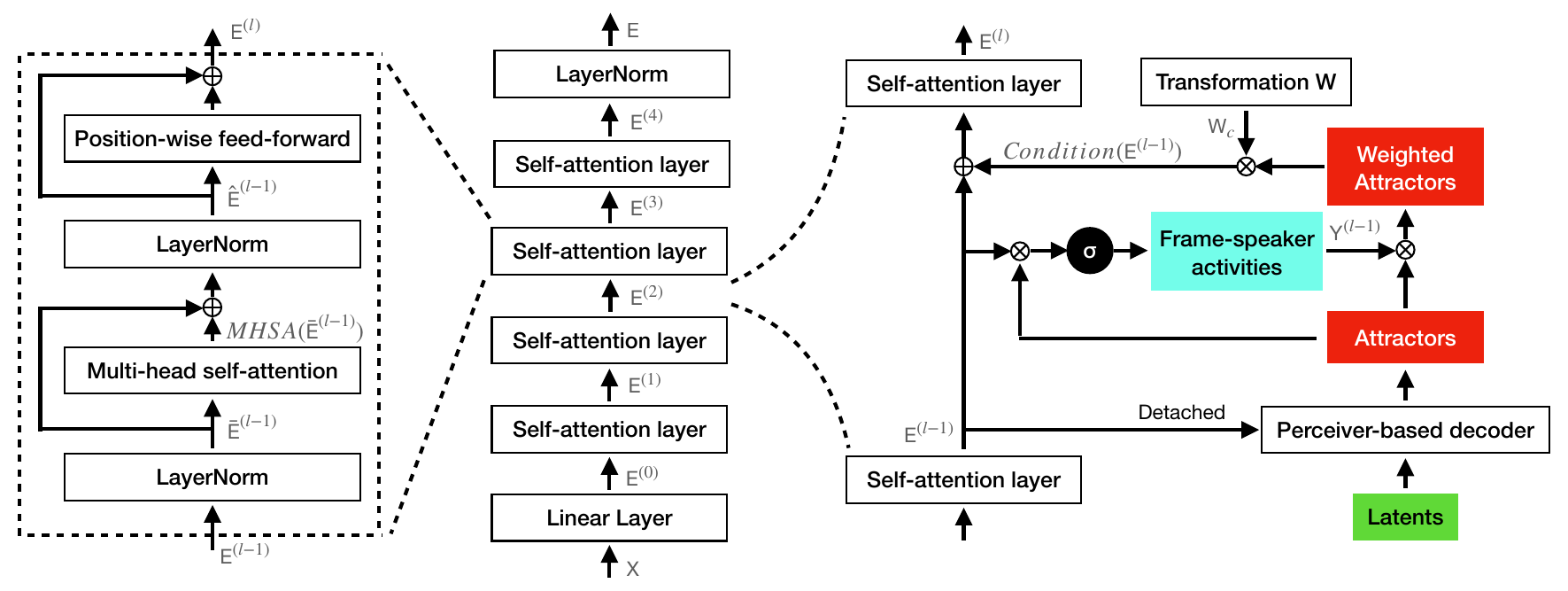}
\vspace{-7mm}
\caption{Scheme of frame encoder (middle), detail of self-attention layer (left) and conditioning scheme (right).}
\label{fig:Frame_Encoder_extended}
\end{figure*}

The main two modules in DiaPer are the frame encoder and the attractor decoder. As shown in Figure~\ref{fig:Frame_Encoder_extended} and proposed in~\cite{fujita2019end}, the frame encoder receives the sequence of frame features $\textbf{X}$ and transforms them with a few chained self-attention layers $\textbf{E} = FrameEncoder(\textbf{X})$ to obtain the frame embeddings $\textbf{E} \in \mathbb{R}^{T \times D}$. The attractor decoder receives the frame embeddings and produces attractors $\textbf{A} = PercDec(\textbf{E})$ with $\textbf{A} \in  \mathbb{R}^{A \times D}$\footnote{In practice, $S=A$.} which are in turn compared with the frame embeddings to determine which speaker is active at each time-frame: $\hat{\textbf{Y}} = \sigma(\textbf{E} PercDec(\textbf{E})^\top)$.

In other words, the frame encoder is in charge of transforming the initial input features into deeper and more contextualized representations from which (a) the attractors will be estimated, and (b) the frame-wise activation of each speaker will be determined. Several encoder layers are used to extract such representations and, in a similar way as presented in~\cite{fujita2023neural}, each layer also includes frame-speaker activities conditioning. As shown in Figure~\ref{fig:Frame_Encoder_extended}, intermediate attractors are calculated given the frame embeddings of each frame encoder layer. The intermediate attractors are then weighted by intermediate frame activities and transformed into the frame embedding space to produce the conditioning.
While in EEND-EDA the input frame embeddings are directly processed to obtain attractors, attractors obtained with an attention mechanism need queries to be compatible with keys. The intermediate loss ensures that they match at different encoder layers, thus easing the compatibility at the end of the frame encoder. The attractors are always calculated with the same Perceiver-based decoder, i.e. the parameters are shared for all the intermediate attractors.

More formally, the $FrameEncoder$ consists of
\begin{flalign}
    &\textbf{e}_t^{(0)} = \textbf{W}_{in} \textbf{x}_t + \textbf{b}_{in}& \\
    &\textbf{E}^{(0)} = [\textbf{e}_1^{(0)}, \ldots, \textbf{e}_T^{(0)}]\\
    &\textbf{E}^{(l)} = FrEncLayer_l(\textbf{E}^{(l-1)} + Condition(\textbf{E}^{(l-1)}))& 
    \label{eq:encoder_plus_condition}
\end{flalign}
where $1 \leq l \leq L$, and $L$ is the number of self-attention layers ($FrEncLayer_l$ denoting the $l^{th}$ self-attention layer) and $\textbf{W}_{in} \in \mathbb{R}^{D \times F}$ and $\textbf{b}_{in} \in \mathbb{R}^{D}$ are the weights and biases of the input transformation on the frames.
\begin{flalign}
    &\bar{\textbf{E}}^{(l-1)} = LN(\textbf{E}^{(l-1)})& \\
    &\hat{\textbf{E}}^{(l-1)} = LN(\bar{\textbf{E}}^{(l-1)} + MHSA^{(l)}(\bar{\textbf{E}}^{(l-1)}))& \\
    &FF(\hat{\textbf{E}}^{(l-1)}) = ReLU(\hat{\textbf{E}}^{(l-1)} \textbf{W}_1^{(l)}  + \mathbf{1} \textbf{b}_1^{(l)\top}) \textbf{W}_2^{(l)} + \mathbf{1} \textbf{b}_2^{(l)\top}& \\
    &\textbf{C}_h^{(l)} = Softmax \Big( \frac{\bar{\textbf{E}}^{(l-1)} \textbf{Q}_h^{(l)} (\bar{\textbf{E}}^{(l-1)} \textbf{K}_h^{(l)})^\top}{\sqrt{d}} \Big) (\bar{\textbf{E}}^{(l-1)} \textbf{V}_h^{(l)}) & \\
    &MHSA^{(l)}(\bar{\textbf{E}}^{(l-1)}) = [ \textbf{C}_1^{(l)} \ldots \textbf{C}_H^{(l)} ] \textbf{O}^{(l)}& \\
    &FrEncLayer(\textbf{E}^{(l-1)}) = \hat{\textbf{E}}^{(l-1)} + FF(\hat{\textbf{E}}^{(l-1)}) & 
\end{flalign}
where $H$ is the number of heads (with $1 \leq h \leq H$), $\textbf{W}_1 \in \mathbb{R}^{D \times D_{ff}}$, $\textbf{W}_2 \in \mathbb{R}^{D_{ff} \times D}$, $\textbf{b}_1 \in \mathbb{R}^{D_{ff}}$, $\textbf{b}_2 \in \mathbb{R}^{D}$ are the weights and biases of the position-wise feed-forward layer, $\mathbf{1} \in \mathbb{R}^{T}$ is an all-one vector, $ReLU(\cdot)$ is the rectified linear unit activation function, $\textbf{Q}_h^{(l)} \in \mathbb{R}^{D \times d}$, $\textbf{K}_h^{(l)} \in \mathbb{R}^{D \times d}$, $\textbf{V}_h^{(l)} \in \mathbb{R}^{D \times d}$, $\textbf{O}_h^{(l)} \in \mathbb{R}^{D \times D}$ are the query, key, value and output projection matrices for the $h$\textsuperscript{th} head and $l$\textsuperscript{th} layer, and $d = \frac{D}{H}$ is the dimension of each head. LN stands for layer normalization, MHSA stands for multi-head self-attention and, FF stands for feed-forward layer.

The conditioning is defined as follows
\begin{flalign}
    &Condition(\textbf{E}^{(l-1)}) = \hat{\textbf{Y}}^{(l-1)} PercDec(\textbf{E}^{(l-1)}) \textbf{W}_c& \\
    &\hat{\textbf{Y}}^{(l-1)} = \sigma(\textbf{E}^{(l-1)} PercDec(\textbf{E}^{(l-1)})^\top),&
\end{flalign}
where $PercDec$ is the Perceiver-based attractor decoder, $\textbf{W}_c \in \mathbb{R}^{D \times D}$ is a learnable parameter that weights the effect of the intermediate attractors on the frame embeddings. The application of the conditioning mechanism allows the intermediate frame embeddings to be contextualized given the attractors\footnote{This mechanism could also be understood as a cross-attention operation where the frame embeddings function as queries and the attractors as keys and values, and the attention weights are given by the frame-speaker activities.}. We utilize the term ``conditioning'' to be consistent with~\cite{fujita2023neural}.

\begin{figure}[!t]
\centering
\includegraphics[width=0.8\linewidth]{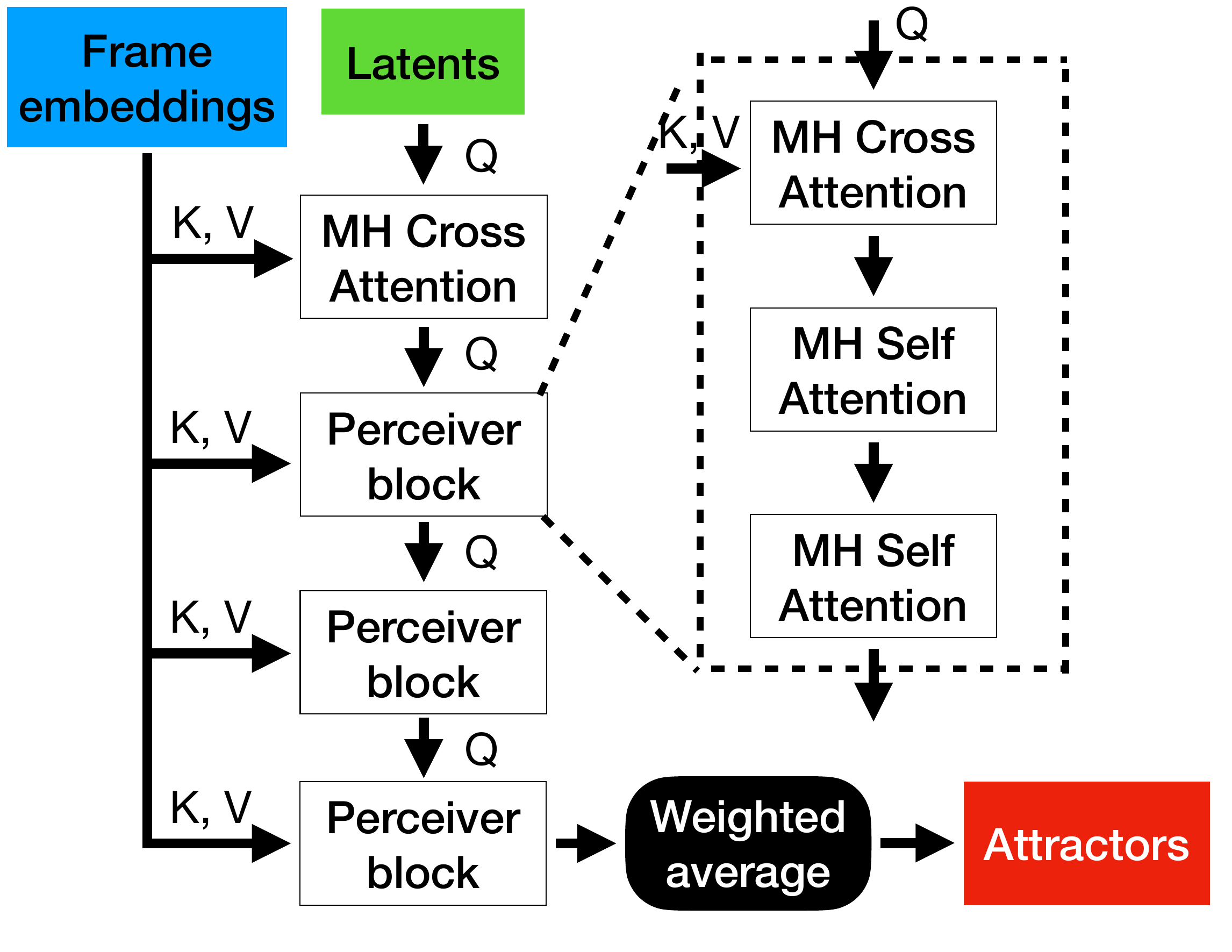}
\vspace{-4mm}
\caption{Scheme of Perceiver decoder.}
\label{fig:perceiver_decoder}
\end{figure}

The decoder makes use of a chain of a few Perceiver blocks as depicted in Figure~\ref{fig:perceiver_decoder}. The set of learnable latents is transformed by each block utilizing the frame embeddings as keys and values. The latents are randomly initialized before starting training and learned during that process. Then, they are fixed at inference time and transformed with the successive application of the Perceiver blocks to be adapted for the given recording.

One could have an equal number of latents and attractors, in which case the latents are an initial representation transformed by the blocks to obtain the attractors. In practice, we observed that this leads to instability in the training and that obtaining the attractors as the linear combination of a larger set of (transformed) latents performed better. More formally,

\begin{small}
\begin{flalign}
    &\textbf{L}^{(0)} = MHA^{(0)}(\textbf{L}, \textbf{E}^{(L)}, \textbf{E}^{(L)})& \\
    &\textbf{L}^{(b)} = PercBlock_b(\textbf{L}^{(b-1)}, \textbf{E}^{(L)})& \\
    &\textbf{C}_h^{(b)} = Softmax \Big( \frac{\textbf{L}^{(b-1)} \textbf{Q}_h^{(b)} (\textbf{E}^{(L)} \textbf{K}_h^{(b)})^\top}{\sqrt{d}} \Big) (\textbf{E}^{(L)} \textbf{V}_h^{(b)}) & \\
    &CA^{(b)} = MHA^{(b)}(\textbf{L}^{(b-1)}, \textbf{E}^{(L)}, \textbf{E}^{(L)}) = [ \textbf{C}_1^{(b)} \ldots \textbf{C}_H^{(b)} ] \textbf{O}^{(b)}& \\
    &PercBlock_b(\textbf{L}^{(b-1)}, \textbf{E}^{(L)}) = MHSA^{(b)_1}(MHSA^{(b)_2}(CA^{(b)}))& \\
    &PercDec(\textbf{E}^{(L)}) = \textbf{W} PercBlock_b(\textbf{L}^{(B)}, \textbf{E}^{(L)}),&
\end{flalign}
\end{small}
where $\textbf{L} \in \mathbb{R}^{L \times D}$ is the set of latents, $B$ is the number of Perceiver blocks in the decoder (with $1 \leq b \leq B$), $H$ is the number of heads (with $1 \leq h \leq H$), $\textbf{Q}_h^{(b)} \in \mathbb{R}^{D \times d}$, $\textbf{K}_h^{(b)} \in \mathbb{R}^{D \times d}$, $\textbf{V}_h^{(b)} \in \mathbb{R}^{D \times d}$, $\textbf{O}_h^{(b)} \in \mathbb{R}^{D \times D}$ are the query, key, value and output projection matrices for the $h$\textsuperscript{th} head and $b$\textsuperscript{th} layer, and $d = \frac{D}{H}$ is the dimension of each head. $MHA$ stands for multi-head cross-attention and $\textbf{W} \in \mathbb{R}^{A \times L}$ is the matrix that linearly combines latents to obtain attractors.

DiaPer decodes always the same fixed number of attractors, denoted by $A$. As mentioned above, the attractors are obtained as a linear combination of the latents. Therefore, the original latents are encouraged to represent information about the speakers in a general manner so that these representations can be transformed (through cross- and self-attention) given a particular input sequence in order to capture the characteristics of the speakers in the utterance. Furthermore, in order to encourage the model to utilize all latents, an extra ``entropy term'' $\mathcal{L}_e$ is added to the loss so that the weights that define the linear combination of latents do not become extreme values (i.e. no latent has a very high weight, therefore making all others very small), where 
\begin{equation}
    \label{eq:entropy_loss}
    \mathcal{L}_e = \sum_{a=1}^{A} mean(Softmax(\textbf{w}_a) * \log Softmax(\textbf{w}_a))
\end{equation}
and $\textbf{w}_a \in \mathbb{R}^{L}$ is the row of $\textbf{W}$ corresponding to attractor $a$.

In standard scaled dot-product attention~\cite{vaswani2017attention}, the softmax is applied on the time-axis to normalize the attention weights along the sequence length before multiplying with the values. In Perceiver, cross- and self-attention on the latents are intertwined. We observed slightly better performance if, when doing cross-attention, the softmax was applied to normalize across latents rather than along the sequence length, i.e. each frame embedding is ``probabilistically'' assigned to each latent using weights that sum up to one. This and other decisions are compared in the experimental section.

As usual for EEND-based models, the diarization loss $\mathcal{L}_d$ is calculated as 
\begin{equation}
    \label{eq:diar_loss}
    \hat{\mathcal{L}}_d(\textbf{Y}, \hat{\textbf{Y}}) = \frac{1}{T S} \ \underset{\phi \in perm(S)}{\min} \sum_{t}^T BCE(\mathbf{y}_t^{\phi}, \hat{\mathbf{y}}_t),
\end{equation}
where considering all reference labels permutations denote permutation invariant training (PIT) loss.

Like in EEND-EDA, to determine which attractors are valid, an attractor existence loss $\hat{\mathcal{L}}_a$ is calculated as $\hat{\mathcal{L}}_a(\mathbf{r}, \mathbf{p}) = BCE(\mathbf{r}, \mathbf{p})$ using the same permutation given by $\hat{\mathcal{L}}_d$.

$\hat{\mathcal{L}}_d$ and $\hat{\mathcal{L}}_a$ are enough to train the model, but inspired by other works~\cite{yu2022auxiliary,fujita2023neural,jeoung2023improving}, we decided to introduce auxiliary losses.
The main idea is that using the frame embeddings produced by the frame encoder, we calculate losses using the intermediate attractors given by the latents after each Perceiver block. Analogously, using the attractors produced by the Perceiver-based decoder, we calculate losses using the intermediate frame embeddings given after each layer in the frame encoder. The averages of the intermediate losses over frame encoder layers and over Perceiver blocks are summed to the losses $\hat{\mathcal{L}}_d(\textbf{Y}, \hat{\textbf{Y}})$ and $\hat{\mathcal{L}}_a(\mathbf{r}, \mathbf{p})$ which use ``final'' attractors and ``final'' frame embeddings.
Then, $\mathcal{L}_d$ and $\mathcal{L}_a$ are obtained as
\begin{small}
\begin{flalign}
    \mathcal{L}_d = \hat{\mathcal{L}}_d(\textbf{Y}, \hat{\textbf{Y}}) + \frac{1}{L-1} \sum_{l=1}^{L-1} \hat{\mathcal{L}}_d(\textbf{Y}, \hat{\textbf{Y}}^l) + \frac{1}{B-1} \sum_{b=1}^{B-1} \hat{\mathcal{L}}_d(\textbf{Y}, \hat{\textbf{Y}}^b) \\
    \mathcal{L}_a = \hat{\mathcal{L}}_a(\mathbf{r}, \mathbf{p}) + \frac{1}{L-1} \sum_{l=1}^{L-1} \hat{\mathcal{L}}_a(\mathbf{r}, \mathbf{p}^l) + \frac{1}{B-1} \sum_{b=1}^{B-1} \hat{\mathcal{L}}_a(\mathbf{r}, \mathbf{p}^b),
\end{flalign}
\end{small}
where $\mathbf{p} = [ p_1, \ldots, p_{A} ]$ are the attractor posterior existence probabilities and $\mathbf{r} = [ r_1, \ldots, r_{A} ]$ are the reference presence labels $r_i \in \{0, 1\}$ for $1 \leq i \leq A$. $\mathbf{p}^l$ are the posteriors using the frame embeddings of the $l$\textsuperscript{th} frame encoder layer and $\mathbf{p}^b$ are the posteriors using the $b$\textsuperscript{th} Perceiver block.

The final loss to be optimized is $\mathcal{L} = \mathcal{L}_d + \mathcal{L}_a + \mathcal{L}_e$.

One of the major disadvantages when using a non-autoregressive decoder is that the number of elements to decode (attractors in this case) has to be set in advance and this imposes a limit on the architecture. However, unlike the original versions of EEND, we do not focus on a scenario with a specific quantity of speakers but rather set the model to have a maximum number of attractors $A$ large enough to handle several scenarios. This is done in one way or another in all methods that handle ``flexible'' amounts of speakers, i.e. when running inference with EEND-EDA, it is necessary to decode a specific maximum number of attractors. DiaPer decodes always the same number of attractors and, like in EEND-EDA~\cite{horiguchi20_interspeech}, a linear layer plus sigmoid determine which attractors are valid, i.e. correspond to a speaker in the conversation.

\section{Experimental Setup}
\subsection{Data}
\subsubsection{Training data}
One of the key aspects of training end-to-end diarization models is the training data. Neural models require large amounts of training data annotated for diarization which, in practice, are scarce. The compromise solution consists in generating training data artificially by combining segments of speech from different recordings. Simulated mixtures~\cite{fujita19_interspeech} have been shown to enable the training of EEND models but they have some disadvantages, mainly related to their lack of naturalness. Some works~\cite{yamashita22_odyssey,landini22_interspeech,landini2023multi} have explored alternatives that allow these models to obtain better performance. In this work, we opt for simulated conversations (SC) for which public recipes are available\footnote{\url{https://github.com/BUTSpeechFIT/EEND\_dataprep}} and for which the advantages over mixtures have been shown for real conversations with two and more speakers~\cite{landini22_interspeech,landini2023multi}.

Following this approach, different sets of SC were generated. To train 8\,kHz models, 10 sets were created, each with a different number of speakers per SC (ranging from 1 to 10) and each containing 2500\,h of audio. Utterances from the following sets were used: Switchboard-2 (phases I, II, III)~\cite{graff1998switchboard,graff1999switchboard,graff2002switchboard}, Switchboard Cellular (parts 1 and 2)~\cite{graff2001switchboard,graff2004switchboard}, and NIST Speaker Recognition Evaluation datasets (from years 2004, 2005, 2006, 2008)~\cite{nist20062004,nist20112005,nist20112005test,nist20112006test,nist20112006train,nist20122006test,nist20112008train,nist20112008test}. All the recordings are sampled at 8\,kHz and, out of 6381 speakers, 90\% are used for creating training data.
The Kaldi ASpIRE VAD\footnote{\url{http://kaldi-asr.org/models/m4}} is used to obtain time annotations (in turn used to produce reference diarization labels).
To augment the training data, we use 37 noises from MUSAN~\cite{snyder2015musan} labeled as ``background''. They are added to the signal scaled with a signal-to-noise ratio selected randomly from \{5, 10, 15, 20\} dB.

In order to train 16\,kHz models, a similar strategy was followed to also generate SC with different amounts of speakers ranging from 1 to 10 per conversation, all comprised of 2500\,h of audio. Instead of telephone conversations, utterances were taken from LibriSpeech~\cite{panayotov2015librispeech} which consists of 1000 hours of read English speech from almost 2500 speakers. The same VAD as described above was used to produce annotations and equivalent background noises were used, but in 16\,kHz.

\subsubsection{Evaluation data}
\label{sec:evaldata}
Different corpora were used to evaluate the models. For telephone speech, we utilized the speaker segmentation data from 2000 NIST Speaker Recognition Evaluation~\cite{przybocki2001nist} dataset, usually referred to as ``Callhome''~\cite{NISTSRE2000evalplan} which has become the de facto telephone conversations evaluation set for diarization containing recordings with different numbers of speakers as shown in Table~\ref{tab:callhome_speakers}. 
We report results using the standard Callhome partition\footnote{Sets listed in \url{https://github.com/BUTSpeechFIT/CALLHOME\_sublists}}, denoting the partitions as CH1 and CH2. We also report results on the subset of 2-speaker conversations to which we refer as CH1-2spk and CH2-2spk.
Results on Callhome consider all speech (including overlap segments) for evaluation with a forgiveness collar of 0.25\,s. 
We also report results on the conversational telephone speech (CTS) domain from the Third DIHARD Challenge~\cite{ryant21_interspeech}, which consists of previously unpublished telephone conversations from the Fisher collection. The development and evaluation sets in the ``full'' set consist of 61 2-speaker 10-minute recordings each. Originally 8\,kHz signals, they were upsampled to 16\,kHz for the challenge and downsampled to 8\,kHz to be used in this work. As usual on DIHARD, all speech is evaluated with a collar of 0\,s.

Besides telephone conversations, we compared the models on a variety of wide-band datasets. As the models we evaluate are trained on single-channel data, when the datasets contain microphone array data, we mix all channels in the microphone array (far-field) or headsets (near-field)\footnote{It should be noted that other works in Table~\ref{tab:multi_wideband_new} might have processed the channels differently so the acoustic inputs for the recordings might differ.}. Training sets (or development, if train sets are not available) are utilized for fine-tuning. The databases considered are:

\begin{itemize}[leftmargin=0.4cm]
    \item AISHELL-4~\cite{fu21b_interspeech}, using the train/evaluation split provided.
    \item AliMeeting~\cite{yu2022m2met}, using the train/eval/test split provided. Unlike in the M2MET Challenge, oracle VAD is not used.
    \item AMI~\cite{carletta2005ami,kraaij2005ami}, using the full-corpus-ASR partition into train/dev/test and the diarization annotations of the ``only words'' setup described in~\cite{landini2022bayesian}\footnote{\url{https://github.com/BUTSpeechFIT/AMI-diarization-setup}}.
    \item CHiME6~\cite{watanabe20b_chime}, using the official partition and annotations from CHiME7 challenge~\cite{cornell23_chime} into train/dev/eval. 
    \item DIHARD2~\cite{ryant2019second}, using the official partition.
    \item DIHARD3~\cite{ryant21_interspeech}, using the official ``full'' partition in order to have a more distinct corpus wrt DIHARD 2. 
    \item DipCo~\cite{segbroeck20_interspeech}, using the official partition and annotations from CHiME7 challenge~\cite{cornell23_chime} into dev/eval. 
    \item Mixer6~\cite{brandschain2010mixer}, using the official partition and annotations from CHiME7 challenge~\cite{cornell23_chime} into train/dev/eval but, given that the train part has only one speaker per recording, we only consider the dev and eval parts. 
    \item MSDWild~\cite{liu22t_interspeech}, using the official partition into few.train/many.val/few.val as train/dev/test following other works.
    \item RAMC~\cite{yang22h_interspeech}, using the official partition.
    \item VoxConverse~\cite{chung20_interspeech}, using the official partition into dev/test and latest annotations\footnote{Version 0.3 in \url{https://github.com/joonson/voxconverse/tree/master}}.
\end{itemize}
More information about each dataset can be found in Table~\ref{tab:wideband_datsets_information}. The choice of forgiveness collar for calculating DER corresponds to the least forgiving choice (i.e. collar of 0\,s) except in cases where a challenge or the authors proposed differently. In no case is used any kind of oracle information (such as VAD) in order to have full pipeline comparisons.

\begin{table}[t]
    \caption{Information per list for Callhome parts 1 and 2.}
    \label{tab:callhome_speakers}
    \vspace{-2mm}
    \setlength{\tabcolsep}{6pt} 
    \centering
    \begin{tabular}{lcccccc|l}
    \toprule
    No. speakers & 2 & 3 & 4 & 5 & 6 & 7 & \# Hours (2-spk)\\ 
    \midrule
    CH1 & 155 & 61 & 23 & 5 & 3 & 2 & 8.70 (3.19) \\
    CH2 & 148 & 74 & 20 & 5 & 3 & 0 & 8.55 (2.97) \\
    \bottomrule
  \end{tabular}
\end{table}

\begin{table*}[t]
    \caption{Information about the number of files, the minimum and maximum number of speakers per recording and the number of hours per partition as well as evaluation collar, types of microphone and characteristics of each evaluation dataset.}
    \label{tab:wideband_datsets_information}
    \vspace{-2mm}
    \setlength{\tabcolsep}{3pt} 
    \centering
    \begin{tabular}{l|ccc|ccc|ccc|cll}
    \toprule
    \multirow{2}{*}{Dataset} & \multicolumn{3}{c|}{train} & \multicolumn{3}{c|}{development} & \multicolumn{3}{c|}{test} & \multicolumn{1}{c}{DER} & \multirow{2}{*}{Microphone} & \multirow{2}{*}{Characteristics} \\ 
    & \#files & \#spk & \#\,h & \#files & \#spk & \#\,h & \#files & \#spk & \#\,h & collar (s) &  &  \\
    \midrule
   AISHELL-4 & 191 & 3-7 & 107.53 & -- & -- & -- & 20 & 5-7 & 12.72 & 0 & array & Discussions in Mandarin in different rooms \\
   AliMeeting & 209 & 2-4 & 111.36 & 8 & 2-4 & 4.2 & 60 & 2-4 & 10.78 & 0 & array \& headset & Meetings in Mandarin in different rooms \\
   AMI & 136 & 3-5 & 80.67 & 18 & 4 & 9.67 & 16 & 3-4 & 9.06 & 0 & array \& headset & Meetings in English in different rooms \\
   CHiME6 & 14 & 4 & 35.68 & 2 & 4 & 4.46 & 4 & 4 & 10.05 & 0.25 & array & Dinner parties in home environments\\
   DIHARD2 & -- & -- & -- & 192 & 1-10 & 23.81 & 194 & 1-9 & 22.49 & 0 & varied & Wide variety of domains\\
   DIHARD3 full & -- & -- & -- & 254 & 1-10 & 34.15 & 259 & 1-9 & 33.01 & 0 & varied & Wide variety of domains \\
   DipCo & -- & -- & -- & 5 & 4 & 2.73 & 5 & 4 & 2.6 & 0.25 & array & Dinner party sessions in the same room \\
   Mixer6 & 243 & 1 & 183.09 & 59 & 2 & 44.02 & 23 & 2 & 6.02 & 0.25 & varied & Interviews and calls in English \\
   MSDWild & 2476 & 2-7 & 66.1 & 177 & 3-10 & 4.1 & 490 & 2-4 & 9.85 & 0.25 & varied & Videos of daily casual conversations \\
   RAMC & 289 & 2 & 149.65 & 19 & 2 & 9.89 & 43 & 2 & 20.64 & 0 & mobile phone & Phone calls in Mandarin \\
   VoxConverse & -- & -- & -- & 216 & 1-20 & 20.3 & 232 & 1-21 & 43.53 & 0.25 & varied & Wide variety of videos (different languages) \\
    \bottomrule
  \end{tabular}
\end{table*}

\subsection{Models}
As the main baseline for this work, we utilize end-to-end neural diarization with encoder-decoder attractors (EEND-EDA)~\cite{horiguchi20_interspeech} which is the most popular EEND approach that can handle multiple speakers. The architecture used was exactly the same as that described in~\cite{horiguchi20_interspeech} and we used our PyTorch implementation\footnote{\url{https://github.com/BUTSpeechFIT/EEND}}. 15 consecutive frames of 23-dimensional log Mel-filterbanks (computed over 25\,ms every 10\,ms) are stacked to produce 345-dimensional features every 100ms. These are transformed by the frame encoder, comprised of 4 self-attention encoder blocks (with 4 attention heads each) into a sequence of 256-dimensional embeddings. These are then shuffled in time and fed into the LSTM-based encoder-decoder module that decodes attractors, which are deemed as valid if their existence probability is above a certain threshold. A linear layer followed by the sigmoid function is used to obtain speech activity probabilities for each speaker (represented by a valid attractor) at each time step (represented by an embedding). 

Part of the setup for DiaPer is shared with the baseline, namely the input features, the frame encoder configuration (except in experiments where the number of layers was changed), and the mechanism for determining attractor existence.

Following standard practice with EEND models, the training scheme consists in training the model first on synthetic training data and then performing fine-tuning (FT) using a small development set of real data of the same domain as the test set. In the experiments with more than two speakers, a model initially trained on synthetic data with two speakers per recording is adapted to a synthetic set with a variable number of speakers and finally fine-tuned to a development set. 
We explored training directly on a set with a variable number of speakers and, as in our previous work~\cite{landini2023multi}, we observed that for the same training time the model would not reach the same performance as training on 2-speaker SC and then adapting to a variable number of speakers. While this does not mean that both approaches cannot reach the same performance, training on a large set with variable number of speakers is costly. From a practical perspective, training first on a 2-speaker set and then adapting to more speakers results in shorter training times. Nevertheless, other curriculum learning strategies could result in more overall efficient training pipelines for Diaper and all other EEND systems, which is something yet to be explored by the community.

As clustering-based baseline, we utilize a VBx-based~\cite{landini2022bayesian} system in two flavors: 8\,kHz and 16\,kHz. Two VADs were used: Kaldi ASpIRE\footnote{\url{http://kaldi-asr.org/models/m4}} and pyannote's. The best one of the two was chosen for each dataset based on performance on the development set. To handle overlap, the OSD from pyannote~\cite{Bredin2020} is run and the second speakers are assigned heuristically~\cite{otterson2007efficient} (closest in time speaker). For results on AMI, Callhome and DIHARD 2, the hyperparameters of VBx were the same as those used in \cite{landini2022bayesian}. For the other sets, discriminative VBx (DVBx)~\cite{klement2024discriminative} was used to find optimal hyperparameters automatically.

\subsection{Training}

\begin{table*}[t]
    \caption{Information about training steps. ``*'' in the fourth column refers to different numbers of epochs for different experiments. The sixth column refers to the maximum number of epochs but different experiments used different (not always the last) epochs. The warm-up in Noam was always carried out for 200k steps.}
    \label{tab:training_steps}
    \vspace{-2mm}
    \setlength{\tabcolsep}{5pt} 
    \centering
    \begin{tabular}{llcccccc}
    \toprule
    SR & Step & ID & Init. (\# ep.) & Set & \# max. ep. & LR & Related results \\ 
    \midrule
    \multirow{7}{*}{8\,kHz} & \multirow{2}{*}{Training (2-speaker SC)} & \multirow{2}{*}{A} & \multirow{2}{*}{None} & \multirow{2}{*}{SC2 8\,kHz} & \multirow{2}{*}{100} & \multirow{2}{*}{Noam} & Tables~\ref{tab:perceiver_blocks_comparison}, \ref{tab:number_latents_comparison}, \ref{tab:number_layers_frameencoder_comparison}, \ref{tab:latent_dimension_comparison}, \ref{tab:ablation_comparison}, \ref{tab:ch2two_other_methods} \\
     &  &  &  &  &  &  & and Figures~\ref{fig:latents_dimensions_part1}, \ref{fig:telephone_two_speakers} \\
     \cmidrule(lr){2-8}
     & \multirow{2}{*}{Adaptation (multi-speaker SC)} & B & A* & SC2-7 8\,kHz & 75 & Noam & Figures~\ref{fig:telephone_more_speakers}, \ref{fig:confmatrixavg} and Tables~\ref{tab:ch2_per_speaker}, \ref{tab:ch2_der_detail} \\
     &  & C & A* & SC1-10 8\,kHz & 100 & Noam & Figures~\ref{fig:confmatrixavg}, \ref{fig:confmatrixminutes}, Table~\ref{tab:multi_wideband_new} (2) \\
     \cmidrule(lr){2-8}
     & \multirow{3}{*}{Fine-tuning (with in-domain data)} & D & A* & CH1 2 speakers & 20 & Adam $10^{-5}$ & Figure~\ref{fig:telephone_two_speakers} and Table~\ref{tab:ch2two_other_methods} \\
     &  & E & A* & DIHARD 3 CTS dev & 20 & Adam $10^{-4}$ & Figure~\ref{fig:telephone_two_speakers} \\
     &  & F & B & CH1 & 20 & Adam $10^{-4}$ & Figure~\ref{fig:telephone_more_speakers} and Tables~\ref{tab:ch2_per_speaker}, \ref{tab:ch2_der_detail}, \ref{tab:ch2all_other_methods} \\
    \midrule
    \multirow{3}{*}{16\,kHz} & Training (2-speaker SC) & G & None & SC2 16\,kHz & 100 & Noam & None\\
    \cmidrule(lr){2-8}
     & Adaptation (multi-speaker SC) & H & G (90-100) & SC1-10 16\,kHz & 100 & Noam & Table~\ref{tab:multi_wideband_new} (5) and (7) \\
     \cmidrule(lr){2-8}
     & Fine-tuning (with in-domain data) & I & H (90-100) & Various & Various & Adam $10^{-6}$ & Table~\ref{tab:multi_wideband_new} (6) and (8)\\
     & Fine-tuning (with compound set) & J & H (90-100) & Compound & 400 & Adam $10^{-5}$ & Table~\ref{tab:multi_wideband_new} (9), (10) and (11)\\
     & Fine-tuning (with in-domain data) & K & J (390-400) & Various & Various & Adam $10^{-6}$ & Table~\ref{tab:multi_wideband_new} (12)\\
    \bottomrule
  \end{tabular}
\end{table*}

Most trainings were run on a single GPU. The batch size was set to 32 with 200000 minibatch updates of warm-up respectively. Following~\cite{horiguchi20_interspeech}, the Adam optimizer~\cite{kingma2014adam} was used and scheduled with noam~\cite{vaswani2017attention}. For a few trainings with a variable number of speakers where 4 GPUs were used, the batchsize and warm-up steps were adapted accordingly. Other hyperparameters (i.e. dropout, learning rate) can be seen in the training configuration files shared in the repository.

For FT on a development set, the Adam optimizer was used. Both EEND-EDA and DiaPer were fine-tuned with learning rate $10^{-5}$ for Callhome 2 speakers due to the low amount of development data and with $10^{-4}$ for whole Callhome and DIHARD 3 CTS. For all the other datasets, DiaPer was fine-tuned on the train set using learning rate $10^{-6}$ until the performance on the development set stopped improving (or, in case there was no official training set available, FT on the development set til not further improvement on the test set).

During training (with 2-speaker SC), adaptation (with a variable number of speakers SC), and FT (with in-domain data), batches were formed by sequences of 600 Mel-filterbank outputs, corresponding to 1 minute, unless specified otherwise (i.e. the analysis in Section~\ref{sec:wideband}). These sequences are randomly selected from the generated SC\footnote{The acute reader will notice that it might not be possible to see as many as 10 speakers in 1 minute, this is addressed in the experimental section.}. During inference, the full recordings are fed to the network one at a time. In all cases, when evaluating a given epoch, the checkpoints of the previous 10 epochs are averaged to run the inference.

To compare EEND-EDA and DiaPer on equal ground, we train both models for the same number of epochs, evaluate them after regular intervals and choose the best performing on the development set. For comparisons on 2-speaker scenarios of Callhome, each model is trained for 100 epochs on telephony SC. Every 10 epochs, the parameters of the 10 previous checkpoints are averaged and performance is evaluated on CH1-2spk set to determine the best one. The performance of such model is reported on CH2-2spk set and DIHARD3 CTS full eval before and after FT.

When doing adaptation to more speakers for comparison on Callhome, the best performing 2-speaker model as described above is selected as initialization. The adaptation to a SC set with different amounts of speakers per recording is run for 75 epochs. The parameters of 10 models are averaged every 5 epochs and performance is evaluated on CH1 to determine the best one.  The performance of such model is reported on CH2. This model is also used as initialization when doing FT to a development set. To avoid selecting results on the test set, all fine-tunings are run for 20 epochs and the parameters of the last 10 epochs are averaged to produce the final model.

For comparisons on the variety of wide-band sets, three variants of DiaPer are trained. An 8\,kHz model following a similar approach as described above: trained for 100 epochs on SC of 2 speakers created with telephony speech and then adapted to the SC with 1-10 speakers set for 100 epochs. The 16\,kHz is trained in the same manner but using SC generated from LibriSpeech. Two flavors of this ``wide-band DiaPer'' are used, one with 10 attractors and another with 20 attractors to analyze the impact on datasets with several speakers. For the comparisons on wide-band sets, results are also shown without and with FT.

\subsection{Metrics}
Diarization performance is evaluated in terms of diarization error rate (DER) as defined by NIST~\cite{NISTRT} and using dscore\footnote{\url{https://github.com/nryant/dscore}}.
During inference time, the model outputs are thresholded at 0.5 to determine speech activities. For evaluation sets where a forgiveness collar is used when calculating DER, a median filter with window 11 is applied as post-processing over the speech activities. If the forgiveness collar is 0\,s, no filtering is applied and, instead of running the inference with 10 frames subsampling in the frame encoder, 5 frames only are subsampled as this provides a better resolution in the output. However, due to the high memory consumption when processing very long files, for CHiME6 a subsampling of 15 frames had to be used.
To analyze the models' quality in terms of finding the correct number of speakers, confusion matrices for correct/predicted numbers of speakers are presented for SC with 10 recordings for each quantity of speakers from 1 to 10.

\section{Experiments}
\subsection{Selection of parameters}
\label{sec:selection_parameters}
In order to shed some light on the influence of different aspects of the architecture in DiaPer, we present first a comparison of the performance when varying some key elements. We start from the best configuration we found, namely: 3 Perceiver blocks in the attractor decoder, 128 latents, 4 self-attention layers in the frame-encoder and 128-dimensional latents, frame embeddings and attractors. This configuration is marked with a gray background in the comparisons. The models are trained on 2-speaker SC and no FT is applied. We also considered different numbers of attractors: 5, 10, and 20 but the performance was the same for the 2-speaker scenario. All the experiments in Sections~\ref{sec:selection_parameters},\ref{sec:ablation},\ref{sec:2-speaker_telephone},\ref{sec:multi-speaker_telephone} had models with 10 attractors which is an upper bound on the expected number of speakers in a recording.

Table~\ref{tab:perceiver_blocks_comparison} shows the impact of the number of Perceiver blocks in the attractor decoder. Out of the configurations explored, having 3 blocks presents the best performance.

\begin{table}[!tb]
  \centering
  \caption{Comparison on CH1-2spk when varying the number of Perceiver blocks in the attractor decoder.}
  \label{tab:perceiver_blocks_comparison}
  \vspace{-2mm}
  \setlength{\tabcolsep}{4pt} 
  \begin{tabular}{@{}
                  l
                  S[table-format=1.2] 
                  S[table-format=1.2] 
                  S[table-format=1.2] 
                  S[table-format=1.2] 
                  S[table-format=1.2] 
                  @{}}
  \toprule
   \# Blocks & \multicolumn{1}{c}{1} & \multicolumn{1}{c}{2} & \multicolumn{1}{c}{\cellcolor{lightgray}3} & \multicolumn{1}{c}{4} & \multicolumn{1}{c}{5} \\ 
    \midrule
   DER (\%) & 8.27 & 8.41 & \cellcolor{lightgray}7.96 & 8.44 & 8.09 \\
   \# Parameters (M) & 3.1 & 3.7 & \cellcolor{lightgray}4.3 & 4.9 & 5.5\\
  \bottomrule
  \end{tabular}
\end{table}

Table~\ref{tab:number_latents_comparison} shows how the number of latents can affect the performance. Differences are small for all amounts equal to or below 256, even with as few as 8. Nevertheless, given that the number of parameters is very similar for any configuration, we keep 128 latents as having more could ease the task when more speakers appear in a recording.

\begin{table}[!tb]
  \centering
  \caption{Comparison on CH1-2spk when varying the number of latents.}
  \label{tab:number_latents_comparison}
  \vspace{-2mm}
  \setlength{\tabcolsep}{4pt} 
  \begin{tabular}{@{}
                  l
                  S[table-format=1.2]  
                  S[table-format=1.2]
                  S[table-format=1.2]
                  S[table-format=1.2]
                  S[table-format=1.2]
                  S[table-format=1.2]
                  S[table-format=1.2]
                  @{}}
  \toprule
  \# Latents & \multicolumn{1}{c}{8} & \multicolumn{1}{c}{16} & \multicolumn{1}{c}{32} & \multicolumn{1}{c}{64} & \multicolumn{1}{c}{\cellcolor{lightgray} 128} & \multicolumn{1}{c}{256} & \multicolumn{1}{c}{512} \\
  \midrule
  DER (\%) & 8.15 & 8.14 & 8.29 & 8.10 & \cellcolor{lightgray}7.96 & 8.10 & 8.54 \\
  \# Parameters (M) & 4.29 & 4.29 & 4.29 & 4.30 & \cellcolor{lightgray}4.31 & 4.32 & 4.36 \\
  \bottomrule
  \end{tabular}
\end{table}

Table~\ref{tab:number_layers_frameencoder_comparison} presents a comparison when varying the number of layers in the frame encoder. Standard SA-EEND and EEND-EDA use 4 and some works have used 6 layers. In the case of DiaPer, we do not observe large differences in the performance and obtain the best performance with 4.

\begin{table}[!tb]
  \centering
  \caption{Comparison on CH1-2spk when varying the number of layers in frame encoder.}
  \label{tab:number_layers_frameencoder_comparison}
  \vspace{-2mm}
  \setlength{\tabcolsep}{4pt} 
  \begin{tabular}{@{}
                  l
                  S[table-format=1.2]  
                  S[table-format=1.2]  
                  S[table-format=1.2]  
                  S[table-format=1.2] 
                  @{}}
  \toprule
        \# Layers & \multicolumn{1}{c}{3} & \multicolumn{1}{c}{\cellcolor{lightgray}4} & \multicolumn{1}{c}{5} & \multicolumn{1}{c}{6} \\
        \midrule
        DER (\%) & 8.18 & \cellcolor{lightgray}7.96 & 8.33 & 8.31 \\
        \# Parameters (M) & 3.7 & \cellcolor{lightgray}4.3 & 4.9 & 5.5\\
  \bottomrule
  \end{tabular}
\end{table}

\begin{figure}
    \centering
    \includegraphics[width=\linewidth]{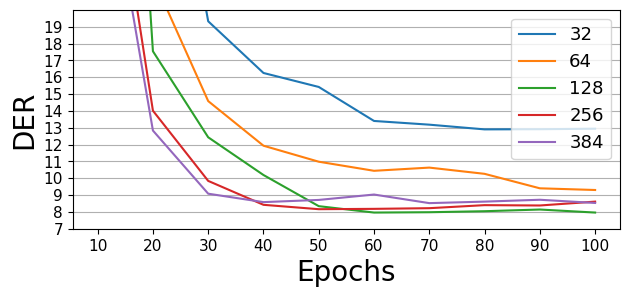}
    \vspace{-4mm}
    \caption{Performance on CH1-2spk for different model dimensions (latents, frame embeddings and attractors).}
    \label{fig:latents_dimensions_part1}
\end{figure}

Finally, Table~\ref{tab:latent_dimension_comparison} shows the impact of the model dimensions on the performance. Increasing the dimensionality of latents, frame embeddings and attractors further than 128 does not show improvements in terms of DER but increases the number of model parameters significantly. Figure~\ref{fig:latents_dimensions_part1} shows performance throughout the epochs for the development set. It is clear how more dimensions allow for a faster convergence; however, more than 128 do not provide more gains in terms of final performance. In addition, more dimensions make the training less stable: using 512 would always lead to instability. Configurations with less than 128 dimensions (64 and 32) can improve further and after 200 epochs reduce the DER by about 1 point but still with worse final results than other configurations. These findings show that reasonable performances can be achieved even with more lightweight versions of DiaPer.

\begin{table}[!tb]
  \centering
  \caption{Comparison on CH1-2spk when varying the model dimension (latents, frame embeddings and attractors).}
  \label{tab:latent_dimension_comparison}
  \vspace{-2mm}
  \setlength{\tabcolsep}{4pt} 
  \begin{tabular}{@{}
                  l
                  S[table-format=2.2]  
                  S[table-format=2.2]  
                  S[table-format=2.2]  
                  S[table-format=2.2]  
                  S[table-format=2.2]
                  @{}}
  \toprule
        Dimensions & \multicolumn{1}{c}{32} & \multicolumn{1}{c}{64} & \multicolumn{1}{c}{\cellcolor{lightgray}128} & \multicolumn{1}{c}{256} & \multicolumn{1}{c}{384}\\ 
    \midrule
        DER (\%) & 12.90 & 9.30 & \cellcolor{lightgray}7.96 & 8.16 & 8.52 \\
        \# Parameters (M) & 0.7 & 1.6 & \cellcolor{lightgray}4.3 & 12.9 & 26.6 \\
  \bottomrule
  \end{tabular}
\end{table}

\subsection{Ablation analysis}
\label{sec:ablation}
Different decisions were made when developing DiaPer and some have a big impact on the performance. Table~\ref{tab:ablation_comparison} presents a comparison of DiaPer in the best configuration shown above and when removing some of the operations performed during training. The first one refers to the normalization of the loss by the reference quantity of speakers, as shown in Eq.~\ref{eq:diar_loss}.
DiaPer always outputs $A$ attractors and the loss is calculated for all of them, even if only training with 2-speakers SC. If the loss is not normalized by the amount of speakers, the model tends to find less speech, increasing the missed speech rate considerably. 

Another ablation is with respect to the frame encoder conditioning described in Figure~\ref{fig:Frame_Encoder_extended}. Similarly to \cite{fujita2023neural}, where the scheme was introduced, removing it worsens the performance by around 0.5 DER. Comparable degradation is observed by removing the loss reinforcements in both frame encoder and Perceiver blocks.

The attention normalization in the cross-attention calculations inside the Perceiver blocks is performed across latents in DiaPer. If done across time, as it is usually done, slightly worsens the performance. We have also explored using across-time normalization in half of the heads and across-latents in the other half but the performance was not better than using across-latents in all heads.

Finally, we also explored removing the entropy loss (Eq.~\ref{eq:entropy_loss}). While the performance is only slightly lower without the loss, the effect might be larger when handling many speakers. The performance on the whole CH1 set was worse in this case, even if the models were trained only with 2-speaker SC.

\begin{table}[!tb]
  \centering
  \caption{DER (\%) on CH1-2spk with different ablation comparisons.}
  \label{tab:ablation_comparison}
  \vspace{-2mm}
  \setlength{\tabcolsep}{4pt} 
  \begin{tabular}{@{}
                  l
                  S[table-format=2.2] 
                  @{}}
  \toprule
  DiaPer & 7.96 \\
  \midrule
  Without normalization of loss per \#speakers & 11.10 \\
  Without frame encoder conditioning & 8.55 \\
  Without intermediate loss in frame encoder & 8.53 \\
  Without intermediate loss in Perceiver blocks & 8.43 \\
  Perceiver cross-attention across time (instead of latents) & 8.07 \\
  Without entropy loss $\mathcal{L}_e$ & 8.02 \\
  \bottomrule
  \end{tabular}
\end{table}

While publications always focus on the positive aspects of the models, we believe there is substantial value in sharing those options that were explored and did not provide gains. Among them were:
\vspace{-0.8mm}
\begin{itemize}[leftmargin=0.4cm]
    \item use absolute positional encoding when feeding the frame embeddings into the attractor decoder (no improvement).
    \item use specaugment for data augmentation (no improvement).
    \item following \cite{maiti2021end,rybicka2022end}, add a speaker recognition loss to reinforce speaker discriminative attractors (slightly worse results).
    \item following \cite{wang2023told}, include an LSTM-based mechanism to model output speaker activities through time (worse performance).
    \item model silence with a specific attractor (worse performance).
    \item Use a linear layer to transform latents into attractors instead of a simple linear combination (learnable) matrix (worse performance).
    \item length normalize frame embeddings and attractors before performing dot-product to effectively compute cosine similarity (worse performance).
    \item use cross-attention to compare frame embeddings and attractors instead of dot-product (worse performance).
    \item as analyzed in \cite{du2021speaker,wang2023told,plaquet23_interspeech}, use power set encoding to model the diarization problem instead of per-frame per-speakers activities (worse performance). In particular, we believe that the reason for this approach not to work with DiaPer is that, when handling many speakers, the number of classes in the power set is too high and most of them are not well represented. This approach has much more potential in limited quantity of speakers scenarios as shown in~\cite{plaquet23_interspeech}.
\end{itemize}
Implementations of most of these variants can be found in our public implementation in \url{https://github.com/BUTSpeechFIT/DiaPer} to enable others to easily revisit them.

\subsection{Two-speaker telephone conversations}
\label{sec:2-speaker_telephone}
\begin{figure}
    \centering
    \subfigure[CH Part 2 (2 speakers)]{{\includegraphics[width=0.48\linewidth]{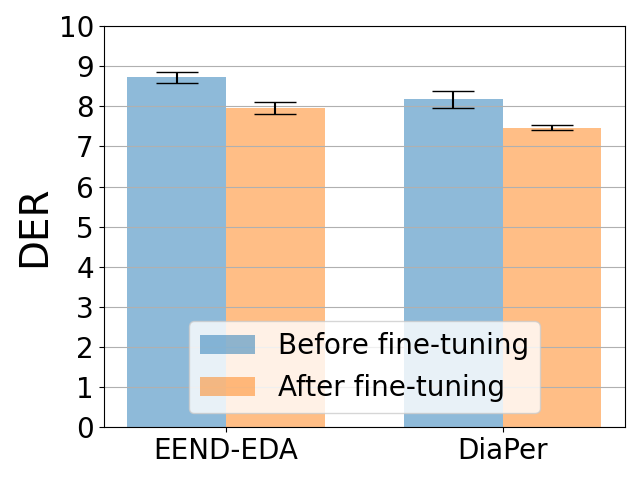} }}%
    \hfill
    \subfigure[DH3 CTS full eval]{{\includegraphics[width=0.48\linewidth]{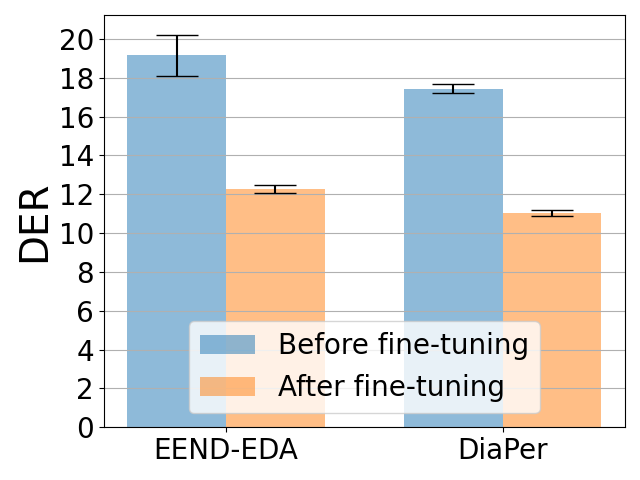} }}
    \vspace{-2mm}
    \caption{DER (\%) for telephone recordings of Callhome and DIHARD 3 conversational telephone speech (CTS) with 2 speakers.}
    \label{fig:telephone_two_speakers}
\end{figure}

Even though DiaPer is specifically designed for the scenario with multiple speakers, as it is common practice, in this section we first present results for the 2-speaker telephone scenario. It should be noted that both EEND-EDA and DiaPer, when trained only with 2-speaker SC learn to only output activities for 2 speakers, even if they are prepared to handle a variable number of them. Figure~\ref{fig:telephone_two_speakers} compares the performance on two sets before and after FT to the in-domain development set. Both EEND-EDA and DiaPer were trained on the same data with 5 different seeds to produce the error bars. Results show that DiaPer can reach significantly better performance on both datasets, both with and without FT.

\begin{figure}
    \centering
    \includegraphics[width=\linewidth]{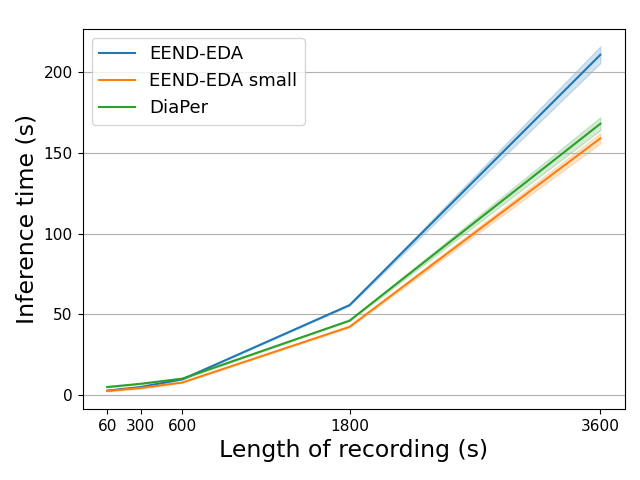}
    \vspace{-5mm}
    \caption{Inference time for EEND-EDA and DiaPer for recordings from 1 minute to 1 hour running 5 times each inference with a downsampling factor of 10. In black is the percentage of time taken by DiaPer wrt EEND-EDA. Ran on Intel(R) Xeon(R) CPU E5-2680 v4 @ 2.40GHz.}
    \label{fig:inferencetime}
\end{figure}

\begin{figure}
    \centering
    \includegraphics[width=\linewidth]{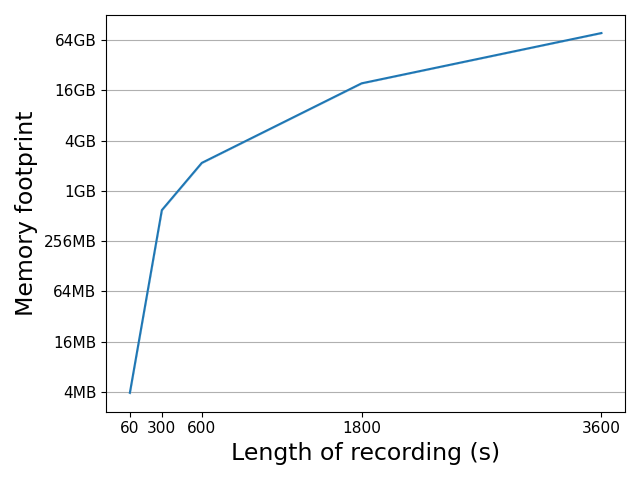}
    \vspace{-5mm}
    \caption{Average inference memory footprint for both EEND-EDA and DiaPer models for recordings from 1 minute to 1 hour running 5 times each inference with a downsampling factor of 10. Experiments ran on Intel(R) Xeon(R) CPU E5-2680 v4 @ 2.40GHz.}
    \label{fig:memory}
\end{figure}

Figure~\ref{fig:inferencetime} presents a comparison between the standard EEND-EDA baseline and DiaPer inference times. Although DiaPer is slower for very short recordings, it can run faster than the standard EEND-EDA when processing several-minute recordings. 
This speed-up is mainly given by the more light-weight nature of DiaPer which results in a faster frame encoder processing, which dominates the computation time versus the attractor decoding in both models. To have a fair comparison, ``EEND-EDA small'' denotes a version of EEND-EDA where the model dimension matches that of DiaPer in its best configuration (128-dimensional frame embeddings and attractors). This corresponds to using the same frame encoder in both models and we can see that DiaPer is slightly slower due to more computations in the attractor decoder. It should be noted that EEND-EDA small performs slightly worse than EEND-EDA in terms of DER. This was not the case for DiaPer and with the smaller configuration, we are able to obtain better DER performance than EEND-EDA and run faster.

Figure~\ref{fig:memory} presents the memory footprint of the models. Note that both EEND-EDA and DiaPer share the same self-attention-based frame encoder architecture which is the main source of memory consumption. Therefore, the memory footprint is a direct function of the sequence length, not the number of parameters of the models. It should be pointed out that these models have very high requirements for long recordings. There is certainly room for improvement regarding this aspect to make end-to-end models more memory efficient.

Table~\ref{tab:ch2two_other_methods} presents an exhaustive comparison with all competitive systems at the time of publication under the same conditions: all speech is evaluated and no oracle information is used. Data refers to the number of hours of data for supervision. For end-to-end models, it can be real or synthetic data and for the clustering-based baseline, it consists of all data used to train the x-vector extractor, VAD and OSD. Methods are divided into groups depending on if they are single or two-stage. Even though DiaPer does not present the best performance among all approaches, it reaches competitive results with fewer parameters and even without FT.

\begin{table}[t]
    \caption[capt]{DER (\%) comparison on CH2-2spk with other methods. For our results, we selected the model with the best performance on CH1 out of the 5 runs. Type can be clustering (C), 1-stage (1-S), or 2-stage (2-S) system. (I) stands for iterative, meaning there is an iterative process at inference time.}
    \label{tab:ch2two_other_methods}
    \vspace{-2mm}
    \setlength{\tabcolsep}{2pt} 
    \centering
    \begin{tabular}{@{}
                  l 
                  c
                  c
                  c
                  c
                  S[table-format=2.2] 
                  S[table-format=2.2]
                  @{}}
    \toprule
    \multirow{2}{*}{System} & \multirow{2}{*}{Type} & \multirow{2}{*}{Code} & \#Param. & Data & \multicolumn{1}{c}{No} & \multicolumn{1}{c}{With} \\
    &  &  & (Million) & (kHour) & \multicolumn{1}{c}{FT} & \multicolumn{1}{c}{FT} \\
    \midrule
    VAD + VBx + OSD & C & \checkmark & 17.9 & 9 & \multicolumn{1}{c}{N/A} & 9.92 \\
    \midrule
    EEND-EDA\cite{horiguchi20_interspeech} & 1-S (I) & \checkmark & 6.4 & 2.4 & \multicolumn{1}{c}{--} & 8.07 \\
    EEND-EDA Confor.\cite{yamashita22_odyssey} & 1-S (I) &  & 4 & 2.5 & 9.65 & 7.18\\
    CB-EEND\cite{liu21j_interspeech} & 1-S &  & 4.2 & 4.7 & \multicolumn{1}{c}{--} & 6.82\\
    DIVE\cite{zeghidour2021dive} & 1-S (I) &  & ?? & 2 & \multicolumn{1}{c}{--} & 6.7 \\
    RX-EEND\cite{yu2022auxiliary} & 1-S &  & 12.8 & 2.4 & \multicolumn{1}{c}{--} & 7.37\\
    EDA-TS-VAD\cite{wang2023target} & 1-S (I) &  & 16.1 & 16 & \multicolumn{1}{c}{--} & 7.04 \\
    EEND-OLA\cite{wang2023told} & 1-S &  & $\approx$6.7 & 15.5 & \multicolumn{1}{c}{--} & 6.91 \\
    EEND-NA\cite{fujita2023neural} & 1-S &  & 5.7 & 2.5 & 8.81 & 7.77\\
    EEND-NA-deep\cite{fujita2023neural} & 1-S &  & 10.9 & 2.5 & 8.52 & 7.12\\
    EEND-IAAE\cite{hao2023nn} (it=2) & 1-S (I) & \checkmark & 8.5 & 2.5 & 13.8 & 7.58 \\
    EEND-IAAE\cite{hao2023nn} (it=5) & 1-S (I) & \checkmark & 8.5 & 2.5 & \multicolumn{1}{c}{--} & 7.36 \\
    AED-EEND\cite{chen23n_interspeech} & 1-S (I) &  & 11.6 & 2.4 & \multicolumn{1}{c}{--} & 6.79 \\
    AED-EEND-EE\cite{chen2024attention} & 1-S (I) &  & 11.6 & 24.7 & \multicolumn{1}{c}{--} & 5.69 \\
    \midrule
    EEND-VC\cite{kinoshita21_interspeech}  & 2-S &  & $\approx$8 & 4.2 & \multicolumn{1}{c}{--} & 7.18 \\
    WavLM + EEND-VC\cite{chen2022wavlm} & 2-S & \checkmark & $\approx$840 & 8 & \multicolumn{1}{c}{--} & 6.46 \\
    EEND-NAA\cite{rybicka2022end} & 2-S (I) &  & 8 & 2.4 & \multicolumn{1}{c}{--} & 7.83\\
    Graph-PIT-EEND-VC\cite{kinoshita22_interspeech} & 2-S &  & $\approx$5.5 & 5.5 & \multicolumn{1}{c}{--} & 7.1\\
    EEND-OLA + SOAP\cite{wang2023told} &  2-S & \checkmark & 15.6 & 19.4 & \multicolumn{1}{c}{--} & 5.73 \\
    \midrule
    \midrule
    EEND-EDA & 1-S (I) & \checkmark & 6.4 & 2.5 & 8.77 & 7.96 \\
    DiaPer & 1-S & \checkmark & 4.6 & 2.5 & 8.05 & 7.51\tablefootnote{It is worth mentioning that out of the 5 runs, the best DER on Part 2 was 7.38 but that did not correspond to the lowest DER on Part 1. Analogously, for EEND-EDA it was 7.78.} \\
    \bottomrule
  \end{tabular}
\end{table}

\subsection{Multiple-speakers telephone conversations}
\label{sec:multi-speaker_telephone}
\begin{figure}
    \centering
    \includegraphics[width=0.65\linewidth]{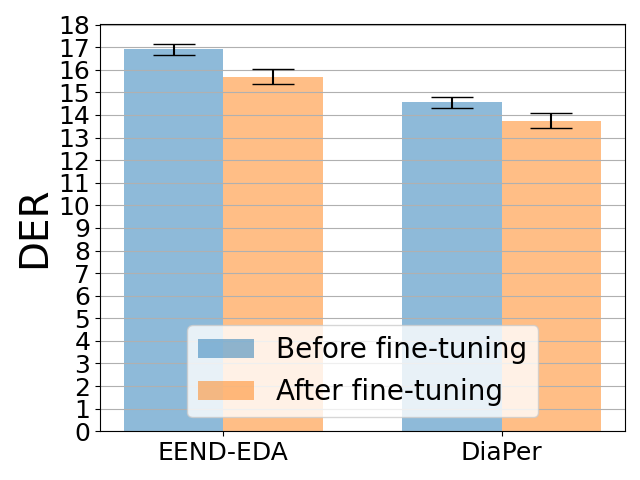}
    \vspace{-4mm}
    \caption{DER (\%) for CH Part 2 with varying number of speakers.}
    \label{fig:telephone_more_speakers}
\end{figure}

Figure~\ref{fig:telephone_more_speakers} presents the comparison for recordings with multiple amounts of speakers where EEND-EDA and DiaPer are trained on the same data. Once again, DiaPer presents significant advantages over EEND-EDA both before and after fine-tuning to the development set. Table~\ref{tab:ch2_per_speaker} shows the DER for different numbers of speakers per conversation where gains are observed in almost all cases. The largest differences are for recordings with more speakers, suggesting the superiority of DiaPer in handling such situations. However, it should be noted that there are only 3 files with 6 speakers and improvements in only one file can affect the results considerably, as is the case here.

\begin{table}[t]
    \caption{DER (\%) comparison on CH2. For each method (EEND-EDA and DiaPer), selecting the best model on CH1 out of the 5 runs.}
    \label{tab:ch2_per_speaker}
    \vspace{-2mm}
    \setlength{\tabcolsep}{5pt} 
    \centering
    \begin{tabular}{@{}
                  l |
                  S[table-format=2.2] |
                  S[table-format=2.2]
                  S[table-format=2.2]
                  S[table-format=2.2] 
                  S[table-format=2.2]
                  S[table-format=2.2] 
                  @{}}
    \toprule
    System & All & \multicolumn{1}{c}{2-spk} & \multicolumn{1}{c}{3-spk} & \multicolumn{1}{c}{4-spk} & \multicolumn{1}{c}{5-spk} & \multicolumn{1}{c}{6-spk} \\
    \midrule
    EEND-EDA & 16.70 & 8.99 & 13.84 & 24.57 & 33.10 & 46.25 \\
    \hspace{0.25cm}+ FT CH1 & 15.29 & 7.54 & 14.01 & 20.84 & 33.34 & 41.36 \\ 
    \midrule
    
    DiaPer & 14.86 & 9.10 & 12.70 & 19.18 & 29.52 & 41.81 \\
    \hspace{0.25cm}+ FT CH1 & 13.60 & 7.39 & 12.08 & 19.62 & 30.25 & 28.84 \\ 
    \bottomrule
  \end{tabular}
\end{table}

Table~\ref{tab:ch2_der_detail} shows the comparison of DER components. It can be observed that 
without fine-tuning DiaPer does not improve the confusion error of EEND-EDA but rather missed and false alarm (FA) speech. A closer look at the inherent VAD and OSD performances of the two models allows us to see that DiaPer improves considerably the OSD recall with similar OSD precision. Therefore, most of the improvement is related to more accurate overlapped speech detection. Nevertheless, it should be pointed out that precision and recall slightly above 50\% are still very low. There is clearly large room for improving the performance in this aspect.

\begin{table}[t]
    \caption{Comparison on CH2. For each method, selecting the best model on CH1 out of the 5 runs. DER and its three components and precision and recall for VAD and OSD performance.}
    \label{tab:ch2_der_detail}
    \vspace{-2mm}
    \setlength{\tabcolsep}{3.5pt} 
    \centering
    \begin{tabular}{@{}
                  l |
                  S[table-format=2.2] |
                  S[table-format=1.2]
                  S[table-format=1.2]
                  S[table-format=1.2] |
                  S[table-format=2.1]
                  S[table-format=2.1] |
                  S[table-format=2.1]
                  S[table-format=2.1] 
                  @{}}
    \toprule
    \multirow{2}{*}{System} & \multicolumn{1}{c|}{DER} & \multicolumn{1}{c}{Miss} & \multicolumn{1}{c}{FA} & \multicolumn{1}{c|}{Conf.} & \multicolumn{2}{c|}{VAD} & \multicolumn{2}{c}{OSD} \\
     & \multicolumn{1}{c|}{(\%)} & \multicolumn{1}{c}{(\%)} & \multicolumn{1}{c}{(\%)} & \multicolumn{1}{c|}{(\%)} & \multicolumn{1}{c}{P (\%)} & \multicolumn{1}{c|}{R (\%)} & \multicolumn{1}{c}{P (\%)} & \multicolumn{1}{c}{R (\%)} \\
    \midrule
    EEND-EDA & 16.70 & 7.08 & 4.88 & 4.73 & 93.3 & 97.6 & 50.0 & 41.9 \\
    \hspace{0.25cm}+ FT CH1 & 15.29 & 8.24 & 2.61 & 4.44 & 95.8 & 94.5 & 63.8 & 38.3 \\ 
    \midrule
    DiaPer & 14.86 & 6.16 & 3.90 & 4.80 & 93.1 & 98.1 & 51.5 & 52.1 \\
    \hspace{0.25cm}+ FT CH1 & 13.60 & 7.80 & 2.06 & 3.74 & 95.4 & 95.3 & 64.1 & 44.8 \\ 
    \bottomrule
  \end{tabular}
\end{table}

\begin{figure}
    \centering
    \subfigure[EEND-EDA]{{\includegraphics[width=0.41\linewidth]{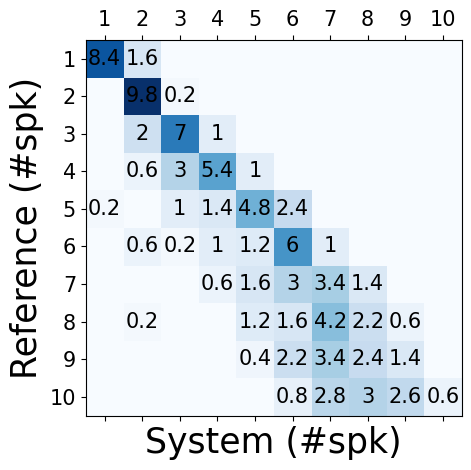} }}%
    \hspace{0.5cm}
    \subfigure[DiaPer]{{\includegraphics[width=0.41\linewidth]{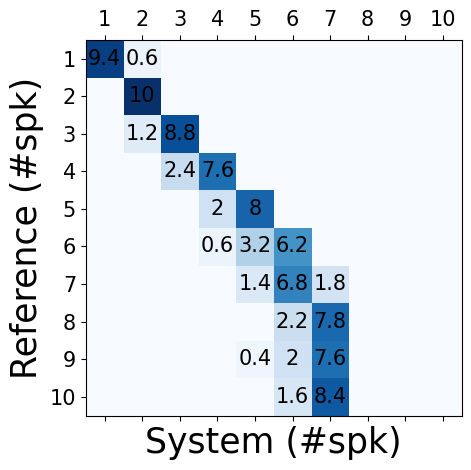} }}%
    
    \vspace{-2mm}
    
    \subfigure[EEND-EDA]{{\includegraphics[width=0.41\linewidth]{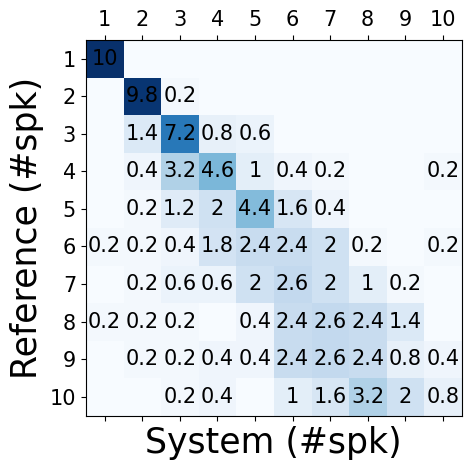} }}%
    \hspace{0.5cm}
    \subfigure[DiaPer]{{\includegraphics[width=0.41\linewidth]{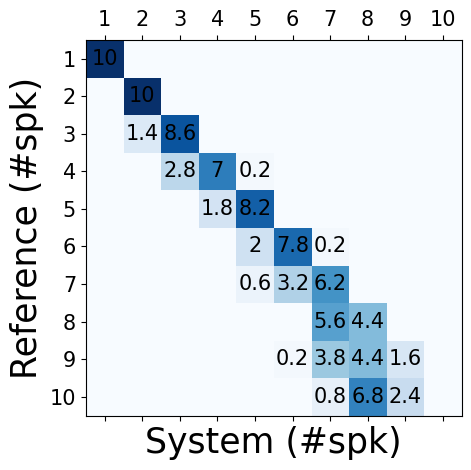} }}%
    \vspace{-2mm}
    \caption{Confusion matrix average of five models evaluated on SC when adapted for 50 epochs with 2-7 speakers (above) and 1-10 speakers (below).}
    \label{fig:confmatrixavg}
\end{figure}

EEND-EDA has been shown to have problems handling several speakers (i.e. not being able to find more than the quantity seen in training and significantly miscalculating the number of speakers when more than 3 are present in a conversation)~\cite{horiguchi2021towards,horiguchi2022encoder}. To compare DiaPer's performance in this sense we trained 5 of both such models with the same procedure and evaluated them on a set of 100 SC with 10 recordings for each number of speakers from 1 to 10. Confusion matrices between the number of real (reference) speakers and the number found by the system were calculated for each model. The averages of such confusion matrices for the 5 DiaPer and 5 EEND-EDA models are presented in Figure~\ref{fig:confmatrixavg}. Although both EEND-EDA and DiaPer are trained on the same data with only up to 7 speakers per SC (matrices above), EEND-EDA is able to find more speakers. Yet, DiaPer is considerably more accurate for SC with up to 6 speakers. When both EEND-EDA and DiaPer are trained with up to 10 speakers per SC (matrices below), we can see that DiaPer is still considerably more accurate. However, its performance is limited when the number of speakers is 8 or more.

\begin{figure}
    \centering
    \subfigure[1 minute, 50 epochs]{{\includegraphics[width=0.41\linewidth]{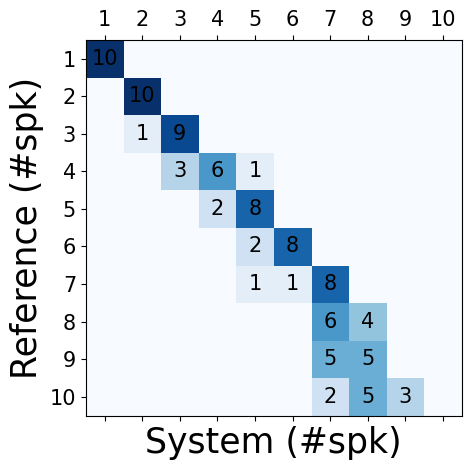} }}%
    \hspace{0.5cm}
    \subfigure[1 minute, 100 epochs]{{\includegraphics[width=0.41\linewidth]{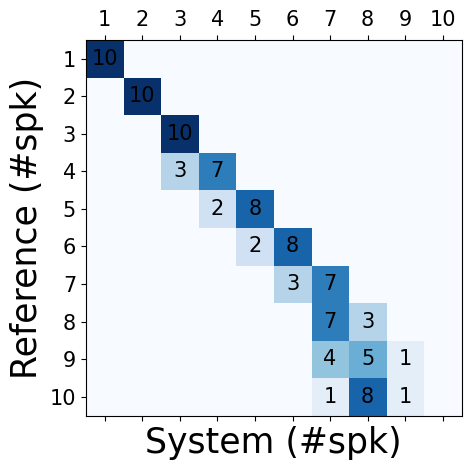} }}%

    \vspace{-2mm}
    
    \subfigure[4 minutes, 50 epochs]{{\includegraphics[width=0.41\linewidth]{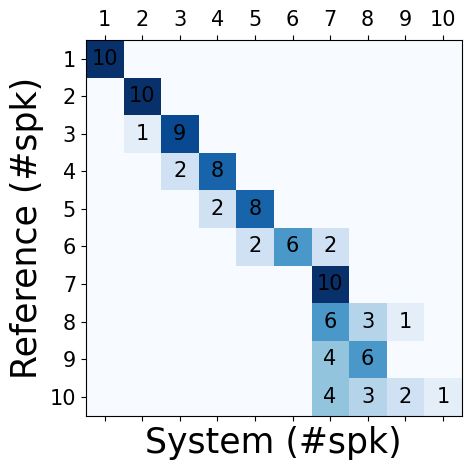} }}%
    \hspace{0.5cm}
    \subfigure[4 minutes, 100 epochs]{{\includegraphics[width=0.41\linewidth]{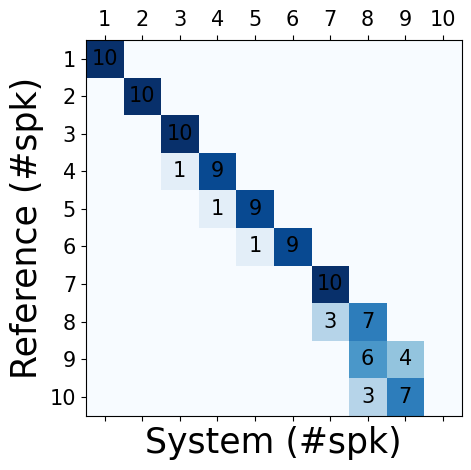} }}%
    \vspace{-2mm}
    \caption{Confusion matrices for DiaPer adapted to telephony SC with 1 to 10 speakers per recording using different sequence lengths to create the batches: 1 minute (top) and 4 minutes (bottom).}
    \label{fig:confmatrixminutes}
\end{figure}

One element to consider is that all the models above were trained and adapted using batches of 1-minute-long sequences. It is less likely for 10 speakers in a simulated conversation to be heard in only one minute. For this reason, we also performed adaptation of one model using 4-minute-long sequences. While sequences of 1 minute have on average 3.6 speakers, sequences of 4 minutes have 5.2, allowing the model to see higher quantities of speakers per training sample during training. A comparison is presented in Figure~\ref{fig:confmatrixminutes} after 50 and 100 epochs training with 1 and 4 minutes sequences. A slight advantage is observed when using 4 minutes after 50 epochs but such advantage increases after 100 epochs.

Finally, Table~\ref{tab:ch2all_other_methods} presents comparisons with other publications on Callhome Part 2 using all recordings. Again, all speech is evaluated and no oracle information is used. For these comparisons, we utilize one of the models trained seeing SC up to 7 speakers (since Callhome does not contain recordings with more speakers). Results show that even if DiaPer has a competitive performance, many methods can reach considerably better results. The main advantage of DiaPer is its lightweight nature, having the least number of parameters in comparison with all other methods. Exploring larger versions of DiaPer (i.e. increasing the model dimension) which could lead to better performance in multi-speaker scenarios is left for future research.

Many previous works present comparisons with clustering-based methods. Although such methods do not deal with overlap intrinsically, it is possible to run an overlapped speech detector and assign second speakers heuristically in order to present a more fair comparison. Interestingly, when utilizing a few years old VAD, VBx and OSD systems, and therefore not highly overtuned, the results are still on par with many end-to-end models showing the relevance of these types of systems even at current time.

\begin{table}[t]
    \caption[capt]{DER comparison on CH2 with other methods. For our results, we selected the model with the best performance on CH1 out of the 5 runs. Type can be clustering (C), 1-stage (1-S) or 2-stage (2-S) system. (I) stands for iterative, meaning there is an iterative process at inference time.}
    \label{tab:ch2all_other_methods}
    \vspace{-2mm}
    \setlength{\tabcolsep}{2pt} 
    \centering
    \begin{tabular}{@{}
                  l 
                  c
                  c
                  c
                  c
                  c
                  S[table-format=2.2] 
                  S[table-format=2.2]
                  @{}}
    \toprule
    \multirow{2}{*}{System} & \multirow{2}{*}{Type} & \multirow{2}{*}{Code} & \#Param. & Data & \multicolumn{1}{c}{No} & \multicolumn{1}{c}{With} \\
    &  &  & (Million) & (kHour) & \multicolumn{1}{c}{FT} & \multicolumn{1}{c}{FT} \\
    \midrule
    VAD + VBx + OSD & C & \checkmark & 17.9 & 9 & \multicolumn{1}{c}{N/A} & 13.63 \\
    \midrule
    EEND-EDA\cite{horiguchi20_interspeech} & 1-S (I) & \checkmark & 6.4 & 15.5 & \multicolumn{1}{c}{--} & 15.29 \\
    EDA-TS-VAD\cite{wang2023target} & 1-S (I) &  & 16.1 & 16 & \multicolumn{1}{c}{--} & 11.18 \\
    EEND-OLA\cite{wang2023told} & 1-S & \checkmark & 6.7 & 15.5 & \multicolumn{1}{c}{--} & 12.57 \\
    AED-EEND\cite{chen23n_interspeech} & 1-S (I) &  & 11.6 & 15.5 & \multicolumn{1}{c}{--} & 14.22 \\
    AED-EEND-EE\cite{chen2024attention} & 1-S (I) &  & 11.6 & 24.7 & \multicolumn{1}{c}{--} & 10.08 \\
    \midrule
    EEND-VC\cite{kinoshita21_interspeech}  & 2-S &  & $\approx$8 & 4.2 & \multicolumn{1}{c}{--} & 12.49 \\
    EEND-GLA\cite{horiguchi2021towards} & 2-S &  & 10.7 & 15.5 & \multicolumn{1}{c}{--} & 11.84 \\
    WavLM + EEND-VC\cite{chen2022wavlm} & 2-S & \checkmark & $\approx$840 & 8 & \multicolumn{1}{c}{--} & 10.35 \\
    Graph-PIT-EEND-VC\cite{kinoshita22_interspeech} & 2-S &  & $\approx$5.5 & 5.5 & \multicolumn{1}{c}{--} & 13.5\\
    EEND-OLA + SOAP\cite{wang2023told} & 2-S & \checkmark & 15.6 & 19.4 & \multicolumn{1}{c}{--} & 10.14 \\
    EEND-VC MS-VBx\cite{delcroix23_interspeech} & 2-S &  & $\approx$840 & 5.5 & \multicolumn{1}{c}{--} & 10.4 \\
    \midrule
    \midrule
    EEND-EDA & 1-S (I) & \checkmark & 6.4 & 15 & 16.70 & 15.29 \\
    DiaPer & 1-S & \checkmark & 4.6 & 15 & 14.86 & 13.60\tablefootnote{It is worth mentioning that out of the 5 runs, the best DER on Part 2 was 13.16 but that did not correspond to the lowest DER on Part 1.}
 \\
    \bottomrule
    \toprule
    \multicolumn{7}{c}{Scoring with collar 0\,s} \\
    \midrule
    VAD + VBx + OSD & C & \checkmark & 17.9 & 9 & \multicolumn{1}{c}{N/A} & 26.18 \\
    \midrule
    pyannote 2.1\cite{bredin23_interspeech} & 2-S & \checkmark & 23.6 & 2.9 & 32.4 & 29.3 \\
    \midrule
    \midrule
    EEND-EDA & 1-S (I) & \checkmark & 6.4 & 2.5 & 28.73 & 25.77 \\
    DiaPer & 1-S & \checkmark & 4.6 & 2.5 & 27.84 & 24.16\tablefootnote{It is worth mentioning that out of the 5 runs, the best DER on Part 2 was 23.81 but that did not correspond to the lowest DER on Part 1.} \\
    \bottomrule
    \end{tabular}
\end{table}

\begin{table*}[t]
    \caption{DER (\%) comparison on a variety of test sets. Overlaps are evaluated and oracle VAD is NOT used. SR stands for sampling rate and ``Cp.'' refers to the compound training set. Results with ``*'' are worse on the test set after fine-tuning but the decision was made on the development set, for which there were improvements. ``Best published results'' refers to the best three reported results at the time of writing. \underline{Underlined} results denote single systems and \textoverline{overlined} results correspond to fusions or more complex models.}
    \label{tab:multi_wideband_new}
    \vspace{-2mm}
    \setlength{\tabcolsep}{3pt} 
    \centering
    \begin{tabular}{@{}
                  clcccccccccccccc
                  @{}}
    \toprule
    ID & System & \rotatebox[origin=c]{90}{SR (kHz)} & \hspace{0.4cm}\rotatebox[origin=c]{90}{\centering AISHELL-4} & \rotatebox[origin=c]{90}{{\centering AliMeeting far}} & \rotatebox[origin=c]{90}{{\centering AliMeeting near}} & \hspace{0.3cm}\rotatebox[origin=c]{90}{\parbox[c]{1.5cm}{\centering AMI array}} & \hspace{0.3cm}\rotatebox[origin=c]{90}{\parbox[c]{1.5cm}{\centering AMI headset}} & \hspace{0.3cm}\rotatebox[origin=c]{90}{\parbox[c]{1cm}{\centering CHiME6}} & \hspace{0.4cm}\rotatebox[origin=c]{90}{DIHARD 2} & \hspace{0.3cm}\rotatebox[origin=c]{90}{{\centering DIHARD 3 full}} & \hspace{0.4cm}\rotatebox[origin=c]{90}{DipCo} & \hspace{0.4cm}\rotatebox[origin=c]{90}{Mixer6} & \hspace{0.4cm}\rotatebox[origin=c]{90}{MSDWild} & \hspace{0.4cm}\rotatebox[origin=c]{90}{RAMC} & \hspace{0.4cm}\rotatebox[origin=c]{90}{VoxConverse} \\
    \midrule
    (1) & VAD+VBx+OSD & 8 & 14.5 & 29.4 & 22.7 & 34.1 & 22.2 & 84.0 & 27.9 & 20.5 & 56.2 & 38.1 & 18.8 & 18.3 & 6.7 \\
    (2) & DiaPer (10att) & 8 & 49.3 & 45.4 & 33.7 & 54.7 & 41.3 & 78.5 & 49.9 & 38.2 & 64.3 & 19.1 & 34.6 & 32.6 & 32.4 \\
    (3) & DiaPer+FT & 8 & 42.7 & 31.6 & 28.9 & 50.5 & 36.4 & 68.5 & 34.1 & 24.1 & 45.1 & 14.7 & 18.0 & 20.9 & 31.6 \\
    \midrule
    (4) & VAD+VBx+OSD & 16 & 15.8 & 28.8 & 22.6 & 34.6 & 22.4 & 70.4 & 26.7 & 20.3 & 49.2 & 35.6 & 16.9 & 18.2 & 6.1 \\
    (5) & DiaPer (10att) & 16 & 48.2 & 38.7 & 28.2 & 57.1 & 36.4 & 78.3 & 43.8 & 34.2 & 48.3 & 21.0 & 35.7 & 38.1 & 23.2 \\
    (6) & (5) + FT 1m & 16 & 41.4 & 32.6 & 27.8 & 49.8 & 32.9 & 70.8 & 33.0 & 24.1 & Overfit & 13.4 & 15.5 & 21.1 & Overfit \\
    (7) & DiaPer (20att) & 16 & 47.9 & 34.4 & 23.9 & 52.3 & 35.1 & 77.5 & 44.5 & 34.8 & 43.4 & 18.5 & 25.1 & 32.1 & 22.1 \\
    (8) & (7) + FT 1m & 16 & 31.3 & 26.3 & 24.4* & 51.0 & 30.5 & 69.9 & 31.2 & 22.8 & Overfit & 11.0 & 14.6 & 18.7 & Overfit \\
    \midrule
    (9) & (7) + FT Cp. 1m & 16 & 39.5 & 27.1 & 23.4 & 53.1 & 33.0 & 67.7 & 34.1 & 25.7 & 44.7 & 14.9 & 14.9 & 21.7 & 26.8 \\
    (10) & (7) + FT Cp. 5m & 16 & 29.4 & 20.7 & 18.2 & 45.3 & 23.9 & 62.6 & 29.1 & 21.1 & 39.6 & 21.9 & 15.3 & 16.2 & 19.6 \\
    (11) & (7) + FT Cp. 10m & 16 & 29.0 & 21.3 & 17.8 & 40.7 & 24.6 & 61.1 & 29.9 & 21.8 & 36.1 & 24.0 & 16.0 & 16.1 & 19.1 \\
    (12) & (7-9) + FT & 16 & 28.8 & 20.2 & 17.6 & 37.5 & 29.1* & 61.6* & 27.7 & 20.3 & 33.8 & 12.5 & 13.4 & 15.7 & 18.2 \\
    \midrule
    \midrule
    & \multicolumn{1}{c}{Best} &  & \underline{16.8}\hspace{1sp}\cite{chen22f_interspeech} & \underline{23.8}\hspace{1sp}\cite{bredin23_interspeech} & --- & \underline{22.2}\hspace{1sp}\cite{bredin23_interspeech} & \underline{18.0}\hspace{1sp}\cite{plaquet23_interspeech} & \textoverline{32.5}\hspace{1sp}\cite{kamontt} & \underline{26.4}\hspace{1sp}\cite{delcroix23_interspeech} & \underline{17.3}\hspace{1sp}\cite{he2023ansd} & \textoverline{22.4}\hspace{1sp}\cite{yeiacas} & \textoverline{7.3}\hspace{1sp}\cite{yeiacas} & \underline{22.0}\hspace{1sp}\cite{liu22t_interspeech} & \underline{22.2}\hspace{1sp}\cite{plaquet23_interspeech} & \textoverline{4.0}\hspace{1sp}\cite{baroudipyannote} \\
    & \multicolumn{1}{c}{published} &  & \underline{14.0}\hspace{1sp}\cite{bredin23_interspeech} & \underline{23.3}\hspace{1sp}\cite{plaquet23_interspeech} & --- & \underline{22.0}\hspace{1sp}\cite{plaquet23_interspeech} & \underline{17.0}\hspace{1sp}\cite{he2023ansd} & \textoverline{27.3}\hspace{1sp}\cite{yeiacas} & \underline{26.9}\hspace{1sp}\cite{horiguchi2021end} & \textoverline{16.9}\hspace{1sp}\cite{horiguchi2021hitachi} & \textoverline{22.0}\hspace{1sp}\cite{wan23_chime} & \textoverline{6.1}\hspace{1sp}\cite{wan23_chime} & \underline{33.6}\hspace{1sp}\cite{liu2022ber} & \underline{19.9}\hspace{1sp}\cite{yang22h_interspeech} & \textoverline{4.4}\hspace{1sp}\cite{karamyan2023krisp} \\
    & \multicolumn{1}{c}{results} &  & \underline{13.2}\hspace{1sp}\cite{plaquet23_interspeech} & \underline{23.5}\hspace{1sp}\cite{raj23_interspeech} & --- & \underline{19.5}\hspace{1sp}\cite{he2023ansd} & \underline{13.0}\hspace{1sp}\cite{chen2024attention} & \textoverline{25.1}\hspace{1sp}\cite{wan23_chime} & \underline{24.6}\hspace{1sp}\cite{chen2024attention} & \textoverline{16.8}\hspace{1sp}\cite{he2023ansd} & \textoverline{16.4}\hspace{1sp}\cite{wan23_chime} & \textoverline{5.7}\hspace{1sp}\cite{kamontt} & \underline{16.0}\hspace{1sp}\cite{plaquet23_interspeech} & \underline{14.4}\hspace{1sp}\cite{broughton23_interspeech} & \textoverline{4.4}\hspace{1sp}\cite{wang2024profile} \\
    \bottomrule
    \end{tabular}
\end{table*}

\subsection{Wide-band scenarios}
\label{sec:wideband}
Most works on end-to-end models focus on the telephone scenario and use Callhome (which is a paid dataset) as benchmark. We believe that this is partly because synthetic data (needed for training such models) match this condition quite well. However, there are many wide-band scenarios of interest when performing diarization and only few works have analyzed their systems on a wide variety of them~\cite{bredin23_interspeech,plaquet23_interspeech}. Following this direction, and pursuing a more democratic field, in this section we use DiaPer on a wide variety of corpora (most of which are of public and free access) and show the performance for the same model (before and after FT) across domains. The results are presented in Table~\ref{tab:multi_wideband_new}.

Since most of the scenarios present many speakers per conversation, all DiaPer models were adapted to the set of 1-10 speakers per recording using sequences of 4 minutes. The 8\,kHz model (system (2)) was trained on telephony SC and two 16\,kHz models were used (systems (5) and (7)). Both wide-band models were trained on LibriSpeech-based SC where one model had 10 attractors (like the 8\,kHz model) and another had 20 attractors to allow for more speakers. All models are evaluated without and with FT (systems (3), (6) and (8)). For corpora where a multi-speaker train set is available, the train set is used for FT until no more improvements are observed on the development set. If no train set is available, the dev set is used for FT until the performance on the test set does not improve further. Therefore, results on these latter corpora should be taken with a grain of salt.

Looking at the results, in some cases, there was overfitting when performing FT on the development set (since those sets did not have a train set). In DipCo, this is most likely due to the limited amount of data. In VoxConverse, the distribution of the number of speakers per recording is skewed towards more speakers in the test set and FT on the dev set makes the model find fewer speakers than without FT. Even more, recordings with more speakers are longer, making the overall error higher after FT on the test set. 

In comparison with the best results published at the time of writing, DiaPer performs considerably worse in most of the scenarios. However, it should be noted that in many cases the best results correspond to systems submitted to challenges which usually consist of the fusion of a few carefully tuned models. 
DiaPer, like any end-to-end system, is very sensitive to the type of training data. This is highly noticeable in the high errors before fine-tuning for all far-field scenarios: AISHELL-4, AliMeeting far mix, AMI mix array, CHiME6 and DipCo;  and relatively lower errors for exclusively close-talk scenarios: AliMeeting near mix, AMI mix headset, Mixer6 and in the comparison between DIHARD 2 and DIHARD 3 full where the latter contains a large portion of telephone conversations. All SC (used to train the models) are generated with speech captured from short distances (telephone for the 8\,kHz system and LibriSpeech for the 16\,kHz ones). Using reverberation could improve the situation, but it has not been explored so far in this context.
Not having enough amount of data matching the testing scenario is a strong drawback for the fine-tuning of end-to-end models as observed with DipCo and VoxConverse. Conversely, Mixer6 and RAMC with large amounts of FT data (more than 100\,h each) and relatively simple setups (interviews and phone calls) are among the scenarios with the largest relative improvement given by the FT.
Even if in most cases the performance is not on par with other approaches, DiaPer's final performance is very competitive for MSDWild and RAMC.

Unlike the telephony scenario where less diverse acoustic characteristics combined with plenty of data enables strong performance of EEND systems even without fine-tuning, the story with wide-band datasets is very different. This is in contrast to traditional clustering-based approaches that are quite robust to different acoustic situations present in wide-band scenarios. Current research suggests that EEND systems do not focus on learning speaker voices~\cite{zhang2024end}, but this might be key to make the EEND systems robust to different conditions like speaker recognition systems are. Devising models and training strategies that can bridge the gap is still an open problem. One possibility would be to capitalize on large amounts of real speaker-labeled data like those normally used to train standard clustering-based systems. Another option could be to make use of data with diarization annotations but very different characteristics.

To explore this idea, and following the strategy shown by Plaquet and Bredin~\cite{plaquet23_interspeech}, we pooled the sets from different corpora to generate a compound training set. Yet, as shown in Section~\ref{sec:multi-speaker_telephone}, the length of the sequences used to construct the training batches can have a relevant effect, especially in scenarios with many speakers. For this reason, three sequence lengths were explored: 1, 5 and 10 minutes corresponding to 2.7, 2.9 and 2.9 speakers per sequence on average for the whole training set. Three fine-tunings on this compound set were performed (one for each sequence length) as shown in systems (9), (10) and (11) in Table~\ref{tab:multi_wideband_new}. Then, for the best of the three configurations for each dataset, a final fine-tuning step was performed using only in-domain data starting from the best-performing system for that dataset among (9), (10) and (11). The sequence length used for the final fine-tuning was the same as the one corresponding to the best FT on the compound system.

In general, fine-tuning on the compound set provides gains for most sets, especially when using 5- or 10-minute sequences. This is beneficial, in particular, for DipCo and VoxConverse for which direct fine-tuning (systems (6) and (8)) does not improve the performance but FT on the compound set provides gains and even enables better performance after FT on the corresponding dev set. This strategy also provides large gains for AliMeeting far and near, AMI array and CHiME6 showing that more FT data can be beneficial on the difficult far-field scenarios.

Regarding the sequence length, it is no surprise that MSDWild reaches the best performance with 1-minute-long sequences since files are less than 1.5 minutes long on average. Surprisingly, Mixer6 also sees degradation when using longer sequences. Even more, a direct fine-tuning (system (8)) performs the best on this set. We believe this could be because there are plenty of hours of data in this relatively simple scenario, combined with the fact that the signal is obtained by combining several channels from multiple devices which could create particular conditions not seen in other datasets. Differences between (10) and (11) are in general small with AMI array and DipCo being the exceptions. While both have long recordings and far-field data which could partly explain this behavior (having longer contexts leads to better representations), the pattern is different for other datasets with similar characteristics such as AISHELL-4, AliMeeting far and, to a lesser extent, CHiME6. The impact of the FT sequence lengths was barely explored here and only some conclusions can be drawn. Nevertheless, we believe this should be better analyzed in the future to devise better training strategies adequate for a given type of data.

Handling extremely long recordings poses difficulties for DiaPer as can be seen for CHiME6. The model probably struggles to condense relevant speaker information for the whole recording on the latent space. It should be noted that the speakers in this dataset can move through different rooms and quite often they sound very quietly. DiaPer consistently finds fewer speakers than the expected four, showing that the model struggles to distinguish the voices. Mechanisms to process the input at different levels (local and global) might help addressing these issues; the clear alternative being EEND-VC-like models but modifications in the encoders to handle the input with different contexts could also provide advantages.

Although we tried to shed some light on reasons for certain training strategies, we believe that many aspects need to be explored. The main goal of this comparison was to present a unified framework evaluated across different corpora. More tailored models could be trained if we used SC with specific numbers of speakers per recording (matching the evaluation data). Likewise, the output post-processing (subsampling and median filter) could be adapted for each dataset. This should definitely result in better performance and is left for future work. 

We can also see that even a standard cascaded system can reach competitive results on a few datasets. This shows the importance and relevance of these systems as baselines nowadays even when end-to-end solutions are the most studied in the community.

Regarding the comparison between 8\,kHz and 16\,kHz DiaPers, in most cases, the latter reaches better performance both without and with FT. Even though the 8\,kHz model was trained with more conversational data, this does not provide advantages over the 16\,kHz model trained on LibriSpeech-based SC. However, the effect of FT is in most cases considerably large, reducing the differences between 8\,kHz and 16\,kHz models. Creating synthetic training data that resembles real ones remains an open challenge for most scenarios.

With respect to the number of attractors in the model, we can observe that overall having more of them is beneficial. This is actually not a drawback for DiaPer since the quantity of attractors does not impact severely on the number of parameters or computations. It is left for future work to explore the effect of larger numbers of attractors (i.e. using 40 or 80).

\section{Conclusions}
In this work, we have presented DiaPer, a new variant of EEND models that makes use of Perceivers for modeling speaker attractors. A detailed analysis of the architectural decisions was presented, including ablations. 
In a thorough comparison on telephone conversations, we showed performance gains wrt EEND-EDA, the most widespread end-to-end model that handles multiple speakers. 

We also presented results on several wide-band datasets comparing the performance with a standard cascaded system and with the best-published results at the time of writing. Even though DiaPer attains competitive performance in some domains, it is considerably worse in others. 

Several aspects are left to study in the future such as changes in the frame encoder, where it seems that the self-attention layers have reached a limit and which present the major hardware bottleneck when handling very long recordings. Furthermore, the frame-encoder and Perceiver blocks could be coupled more tightly to improve the quality of representations (frame embeddings and attractors) simultaneously. 

While DiaPer presents a relatively lightweight end-to-end solution, one avenue for yet more compact models could be parameter sharing: some of the blocks in the architecture could have tied parameters in order to obtain similar results with fewer parameters.

Finally, even if some works have appeared in this direction, how to define proper training sets for end-to-end models is still a very under-explored topic and we believe that further analyses are necessary to bridge the gap in performance between narrow-band and wide-band corpora.

With the aim of facilitating reproducible research, we release the code that implements DiaPer as well as models trained on public and free data.

\section*{Acknowledgments}
We thank the members of the diarization sub-group in BUT for valuable suggestions, especially Anna Silnova for feedback on the manuscript, and Dominik Klement for advice to run DVBx. We also thank Marc Delcroix, Zhengyang Chen, Shota Horiguchi, Zhihao Du, and Chin-Yi Cheng for sharing details about the number of parameters/hours of training data in some models (and all other authors for having shared that information in their work already), and Hervé Bredin for feedback on the presentation of wide-band setups. Finally, we also thank the anonymous reviewers for their comments that allowed us to improve the quality of this work.

The work was supported by the Czech Ministry of Interior project No. VJ01010108 "ROZKAZ", Czech National Science Foundation (GACR) project NEUREM3 No. 19-26934X, and Horizon 2020 Marie Sklodowska-Curie grant ESPERANTO, No. 101007666. Computing on IT4I supercomputer was supported by the Czech Ministry of Education, Youth and Sports through the e-INFRA CZ (IDs 90140 and 90254).

\bibliographystyle{IEEEtran}

\bibliography{mybib}

\end{document}